\documentclass[apj]{emulateapj}

\usepackage{txfonts}
\usepackage{graphicx}
\usepackage{amssymb}
\usepackage{epsfig}
\usepackage{natbib}
\usepackage{textcomp}
\usepackage{color}
\usepackage{rotating}

\newcommand{\ebv}{$E_{B-V}$}

\newcommand{\ebvm}{$E_{B-V}^{\rm{M\,31}}$}

\newcommand{\lam}{$\lambda$}
\newcommand{\n}[1]{$N({\rm #1})$}
\newcommand{\cold}[2]{$#1\times10^{#2}$ cm$^{-2}$}

\newcommand{\be}{\begin{equation}}
\newcommand{\ee}{\end{equation}}
\newcommand{\ie}{\emph{i.e.}}
\newcommand{\eg}{\emph{e.g.}}
\newcommand{\kms}{\mbox{km\,\ensuremath{\rm{s}^{-1}}}}
\renewcommand{\ion}[2]{#1\,\textsc{#2}}

\def\pp{$\phantom{-}$}

\shorttitle{A survey of diffuse interstellar bands in M\,31}
\shortauthors{Cordiner et al.}

\begin{document}

\title{A survey of diffuse interstellar bands in the Andromeda galaxy: \\optical spectroscopy of M\,31 OB stars}

\author{Martin A. Cordiner\altaffilmark{1,2}}
\affil{Astrochemistry Laboratory and The Goddard Center for Astrobiology, Mailstop 691, NASA Goddard Space Flight Center, 8800 Greenbelt Road, Greenbelt, MD 20770, USA}
\email{martin.cordiner@nasa.gov}

\author{Nick L. J. Cox\altaffilmark{3}}
\affil{Institute for Astronomy, K.U. Leuven, Celestijnenlaan 200D, bus 2401,
Leuven, Belgium}

\author{Christopher J. Evans}
\affil{UK ATC, Royal Observatory Edinburgh, Blackford Hill, Edinburgh, EH9 3HJ, UK}

\author{Carrie Trundle}
\affil{Astrophysics Research Centre, School of Mathematics and Physics, Queen's University, Belfast, BT7 1NN, UK}

\author{Keith T. Smith and Peter J. Sarre}
\affil{School of Chemistry, The University of Nottingham, University Park, Nottingham, NG7 2RD, UK}

\and

\author{Karl D. Gordon}
\affil{Space Telescope Science Institute, Baltimore, MD 21218, USA}

\altaffiltext{1}{Formerly: Astrophysics Research Centre, School of Mathematics and Physics, Queen's University, Belfast, BT7 1NN, UK}
\altaffiltext{2}{Working under co-operative agreement with: Department of Physics, The Catholic University of America, Washington, DC 20064, USA}
\altaffiltext{3}{Formerly: Herschel Science Centre, European Space Astronomy Centre, ESA, P.O.Box 78, E-28691 Villanueva de la Ca\~nada, Madrid, Spain}


\begin{abstract}
We present the largest sample to-date of intermediate-resolution blue-to-red optical spectra of B-type supergiants in M\,31 and undertake the first survey of diffuse interstellar bands (DIBs) in this galaxy. Spectral classifications, radial velocities and interstellar reddenings are presented for 34 stars in three regions of M\,31. Based on a subset of these stars with foreground-corrected reddening \ebvm~$\geq$~0.05, the strengths of the M\,31 DIBs are analysed with respect to the amount of dust, the ultraviolet radiation field strength and the PAH emission flux. Radial velocities and equivalent widths are given for the \lam5780 and \lam6283 DIBs towards 11 stars. Equivalent widths are also presented for the following DIBs detected in three sightlines in M\,31: \lam\lam4428, 5705, 5780, 5797, 6203, 6269, 6283, 6379, 6613, 6660, and 6993. All of these M\,31 DIB carriers reside in clouds at radial velocities matching those of interstellar \ion{Na}{i} and/or \ion{H}{i}. The relationships between DIB equivalent widths and reddening (\ebvm) are consistent with those observed in the local ISM of the Milky Way.  Many of the observed sightlines show DIB strengths (per unit reddening) which lie at the upper end of the range of Galactic values. DIB strengths per unit reddening are found (with 68\% confidence), to correlate with the interstellar UV radiation field strength. The strongest DIBs are observed where the interstellar UV flux is lowest. The mean Spitzer 8/24~$\mu$m emission ratio in our three fields is slightly lower than that measured in the Milky Way, but we identify no correlation between this ratio and the DIB strengths in M\,31. Interstellar oxygen abundances derived from the spectra of three M\,31 \ion{H}{ii} regions in one of the fields indicate that the average metallicity of the ISM in that region is $12 + \log[{\rm O}/{\rm H}]=8.54\pm0.18$, which is approximately equal to the value in the solar neighbourhood.
\end{abstract}

\maketitle

\section{Introduction}\label{sec:intro}

Over 400 diffuse interstellar bands (DIBs) are now known \citep{hob08,hob09}, but the identity of the carriers has remained a mystery since their discovery almost 90 years ago. DIBs arise predominantly in the diffuse atomic component of the interstellar medium (ISM) \citep{snow06}, and it is debated whether they are attributable to interstellar gas or dust \citep[see the review by][]{sarre06}. Substructure observed in several DIB profiles indicates that the carriers of these DIBs are probably large, stable, UV-resistant gas-phase molecules \citep{sarre95,ehren96}. Likely carrier candidates include organic molecules such as polycyclic aromatic hydrocarbons (PAHs), fullerenes and carbon chains (see \emph{e.g.} \citealt{salama96,ruiter05}).

The study of DIB strengths in extragalactic environments allows analysis of the carrier formation and destruction mechanisms under physical and chemical conditions that may be quite distinct from those typically found in the Milky Way (MW). Galaxies in the Local Group thus provide us with alternative astrochemical laboratories, and opportunities for detailed studies of interstellar chemistry over a broad range of conditions that may be uncommon or nonexistent in the ISM of our Galaxy. Previous extragalactic DIB research has been reviewed by \citet{snow02a}. Recent research in this area has made significant progress using 10\,m-class telescopes. Atoms, molecules and DIBs have been studied in the Small and Large Magellanic Clouds (MCs) using optical absorption spectroscopy (\emph{e.g.} \citealt{ehren02,cox06,cord06,welty06,cox07}). The effects on DIB carrier abundances of the different physical and chemical interstellar conditions were examined, including the higher gas-to-dust ratios, lower metallicities, different-shaped extinction curves and stronger interstellar radiation fields -- identifying in particular the dependence of DIB strengths (per unit reddening) on the interstellar metallicity. Beyond the Local Group, there have been relatively few studies, which have generally been confined to sightlines probed by bright supernovae \citep{rich87,dodo89,steidel90,sollerman05,cox08,tho08} or background quasars (\eg\ \citealt{york06, ellison08, lawton08}).

Desirable attributes of galaxies for detailed optical spectroscopic absorption studies of the ISM include proximity, an abundance of OB-type supergiants, Doppler velocities greater than $\sim$100 \kms\ (to distinguish extragalactic spectroscopic features from those arising in the Milky Way foreground), and a relatively low foreground extinction. Beyond the Magellanic Clouds, the obvious candidates for study are the spiral galaxies \object[M31]{M\,31} and M\,33. However, these galaxies are more than ten times more distant than the MCs, so high-resolution spectra are not readily obtained using current telescopes; surveys of large numbers of stars are only feasible at low-to-intermediate spectral resolution.  The matter in M\,31 spans a range of radial velocities between around 0 and 600~\kms (see Figure \ref{fig:velocity}), so the majority of spectral features are easily separable from those in the Milky Way, even at modest resolving power. However, as we identify in this study, towards the northern edge of M\,31 the Doppler separation of spectral features can be sufficiently small to cause difficulty in separating the MW and M\,31 components.

Recent work using moderate-resolution (Keck DEIMOS) multi-object spectroscopy of early-type supergiants lead to the first detections of DIBs in M\,31 and M\,33 \citep{cor08,cor08b}. Unusually large DIB equivalent widths per unit reddening (\ebv), compared to the Milky Way, were found in both galaxies. The \lam5780 DIB was subsequently found to be strong relative to \ebv\ in a sample of 17 sightlines in the vicinity of the M\,31 OB78 cluster (NGC\,206) \citep{cox08b}. However, these were selected from an observed set of 72 sightlines based on their strong DIBs relative to the spectral noise, so the results cannot be considered to be representative of the typical behaviour of DIBs in M\,31. In addition, the blue spectra from \citet{cox08b} were of very poor quality, preventing the determination of precise stellar classifications and reddenings. In the present study, emphasis was placed on obtaining higher signal-to-noise spectra throughout the optical to facilitate the measurement of DIBs and the derivation of stellar spectral types.

In terms of its average gas-to-dust ratio \citep{nedia00a,bresolin02} and metallicity \citep[\eg][]{trundle02}, M\,31 has more in common with the Milky Way than the Magellanic Clouds. However, there is strong evidence that the ISM in M\,31 is different to that of the MW and the MCs in the following ways: 1) The M\,31 star formation rate is presently about a tenth that of the MW \citep{walter94,kang09}; 2) the interstellar radiation field is weak in the UV and devoid of far-UV radiation (\emph{e.g.} \citealt{cesarsky98,pagani99,montalto09}) -- interstellar dust grains have apparently not been subjected to the same degree of UV-processing as in the MW \citep{cesarsky98}, which may explain the weakness of the mid-IR PAH bands in the centre of M\,31; 3) the \lam2175~\AA\ UV extinction bump has been measured to be weak and narrow \citep{hutch92,bianchi96}; 4) it has a narrow Serkowski polarisation curve \citep{clayton04}, indicating a different grain-size distribution and/or dust-grain composition; and 5) the extinction law is peculiar as shown by the anomalously low average colour-excess ratio $E_{U-B}/E_{B-V}$ \citep{massey95}. Accepting the current theories of dust-grain extinction (reviewed by \citealt{draine03}), the evidence is consistent with a lack of small graphitic dust grains in the ISM of M\,31 -- a conclusion also reached by \citet{xu94}. \citet{meg01,megier05} identified correlations between Galactic DIB strengths and the shape of the UV extinction curve, which were interpreted as resulting from a physical or chemical relationship between the DIB carriers and the small grains responsible for extinction. Thus, it is of interest to examine whether differences in the size distribution of small grains in the M\,31 ISM compared to the MW give rise to any changes in the DIB spectrum.

\citet{montalto09} concluded, using UV \& mid-IR maps of M\,31, that the observed dust emission is best represented by a model with high PAH abundances ($>$3\% of available carbon) and a low UV field. In their scenario, dust is heated predominantly by an older stellar population, in agreement with the low mean intensity of the UV field. The relative lack of UV photons in M\,31 enables examination of the hypothesis that a strong UV radiation field (on the level of that found in the Milky Way ISM; \eg\ \citealt{draine78}), is required for DIB carrier production (see for example, \citealt{herbig95} and \citealt{kendal02}).

In general, DIB equivalent widths ($W$) correlate well with \ebv\ \citep[\eg][]{herbig93}, and may therefore be used to estimate the amount of dust in a sightline. Thus, DIBs are becoming important as a measure of reddening in external galaxies and DLAs \citep[see for example,][]{lawton08,ellison08}, where they may be used to derive the best -- and sometimes the only -- estimate of \ebv. Given the sparsity of DIB observations outside the Milky Way, and the results of \citet{cox06}, \citet{welty06}, \citet{cor08} and \citet{cor08b}, which suggest that extragalactic DIBs may deviate from the MW \ebv\ relationship, an important goal of our research is to examine the relationship between DIB equivalent widths and reddening in M\,31.

In this article we present the largest sample to-date of blue-to-red spectra of early-type supergiants in M\,31. Following the presentation of their spectral classifications and stellar radial velocities, we use these data to undertake the first survey of DIBs and interstellar \ion{Na}{i} absorption lines across the disc of M\,31, and compare the properties and behaviour of the DIBs with those measured in the Milky Way.

\clearpage

\section{Target selection and observations}

\begin{figure}
\centering
\epsfig{file=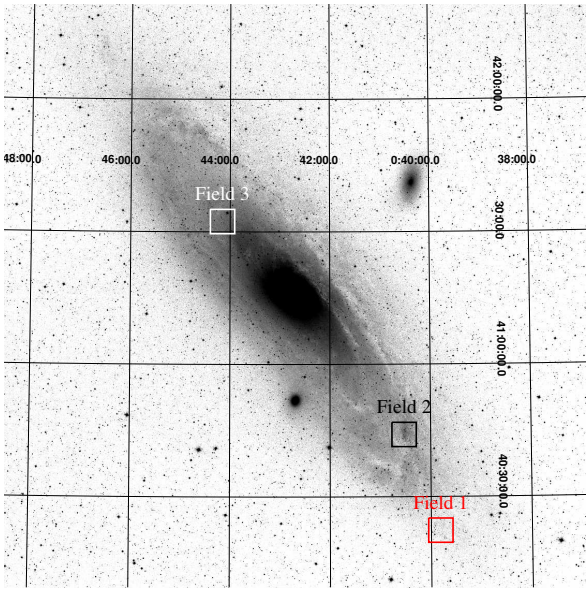, width=0.9\columnwidth}
\caption{Optical image of M\,31 showing observed fields. Individual target locations are plotted in Figure \ref{fig:fields}. Credit: Bill Schoening, Vanessa Harvey/REU program/NOAO/AURA/NSF.}
\label{fig:M31}
\end{figure}

Early-type target stars were selected using $UBVRI$ photometry from the Local Group Galaxies Survey (LGGS) catalogue of 371,782 M\,31 stars \citep{massey06}. Photometric cuts of $V<22$, $-0.45 \leq (B-V) \leq 0.45$, and $-1.2 \leq \rm{Q} \leq -0.4$ (where the reddening-free Johnson $Q$-index is $Q= (U-B) - 0.72(B-V)$), were employed to select the brightest OB supergiants from the catalogue. These cuts revealed a number of OB associations and star clusters across M\,31. To check the accuracy of the colour cuts, stars in each association were cross-matched with published spectral classifications \citep{mac86, massey95} and our own unpublished, low-resolution spectra obtained using the William Herschel Telescope. Three fields were selected for observation using the Gemini Multi-Object Spectrograph (GMOS) on the Gemini North telescope, with the following parameters optimised: 1) the range of galactocentric distances (\ie\ relative to the centre of M\,31); 2) the number of previous OB-type classifications in the region; 3) proximity to H~{\sc ii} regions (the relationship between M\,31 stellar and \ion{H}{ii} region metallicities will be examined in a future article). The three fields observed are in the fields numbered 3, 8, and 9 in the survey of \citet{massey06}. Our fields 2 and 3 (in Massey's Fields 8 and 3), include the OB78 and OB8 clusters, respectively \citep{vandenBergh64}. The three GMOS fields are shown in Figure \ref{fig:M31} and cover regions in M\,31 with galactocentric distances of 14.4-18.4~kpc (Field 1), 8.9-9.6~kpc (Field 2), and $\approx5.8$~kpc (Field 3). The targets in each field were selected to optimise the number of apparently single B-type stars; any stars previously classified as binaries, or which looked particularly asymmetric in the LGGS images, were excluded from the optimisation routine. The orientation of each field was optimised to place the majority of stars in the centre to ensure the optimal spectral coverage for each star. Co-ordinates and photometry for the final selection of 34 target stars are presented in Table \ref{tab:stars}.

\begin{deluxetable*}{lcccclccc}
\tablewidth{0pt}
\tablecaption{Observed target star details \label{tab:stars}}
\tablehead{
Field&Abbr.&LGGS&$V$&$B-V$ &Sp. type &$v_{\rm LSR}$&\ebv&\ebvm\pp\pp\pp
}
\startdata
1 (F9)&3936.51&J003936.51+402235.4&18.538&\pp0.049& B5 Ia & $-520\pm2$&0.13&$0.07\pm0.04$\\
&$\phantom{\star}$~3944.71~$\star$&J003944.71+402056.2&18.200&\pp0.146& O9.7 Ib & $-525\pm5$&0.38&$0.32\pm0.04$\\
&3945.35&J003945.35+402115.4&17.933&\pp0.325& B6 Ia & $-505\pm5$& 0.39&$0.33\pm0.04$\\
&$\phantom{\star}$~3945.82~$\star$&J003945.82+402303.2&18.480&$-$0.064& WN5 $+$ O7 &\nodata &\nodata&\nodata\\
&3954.95&J003954.95+402405.4&18.728&\pp0.014& B8 Ia & $-551\pm2$&0.04&$-0.02\pm0.04$\pp\\
&3956.94$\tablenotemark{t}$&J003956.94+402142.5&18.728&$-$0.137& B0.5 Ia &$-518\pm2$&0.06&$0.00\pm0.04$\\
&3958.22&J003958.22+402329.0&18.968&\pp0.095& B0.7 Ia& $-516\pm2$&0.30&$0.24\pm0.04$\\
&3959.98&J003959.98+402221.1&18.625&$-$0.117& O9.7 Ib & $-514\pm4$&0.11&$0.05\pm0.04$\\
\\
2 (OB78)&4026.74&J004026.74+404418.8&18.387&$-$0.071& B5 Ia & $-579\pm2$&0.01&$-0.05\pm0.04$\pp\\
&4027.18&J004027.18+404233.2&18.009&$-$0.050& B3 Ia & $-571\pm2$&0.03&$-0.03\pm0.04$\pp\\
&4029.57&J004029.57+404457.5&18.192&$-$0.020& B8 Ia & $-590\pm2$&$0.01$&$-0.05\pm0.04$\pp\\
&4029.71&J004029.71+404429.8&18.561&$-$0.229& O7-7.5 Iaf & $-651\pm5$&0.08&$0.02\pm0.04$\\
&$\phantom{\star}$~4030.28~$\star$&J004030.28+404233.1&17.357&$-$0.042& B1 Ia &$-579\pm3$&0.15&$0.09\pm0.04$\\
&$\phantom{\star}$~4030.32~$\star$&J004030.32+404404.8&18.950&$-$0.063& B5 Ia &$-578\pm3$&0.02&$-0.04\pm0.04$\pp\\
&4030.84&J004030.84+404348.6&18.827&$-$0.093& B2 Iab &$-566\pm3$&0.07&$0.01\pm0.05$\\
&4030.94&J004030.94+404246.9&18.866&$-$0.147& O9.5 Ib &$-546\pm5$&0.09&$0.09\pm0.05$\\
&$\phantom{\star}$~4031.52~$\star$&J004031.52+404501.9&18.923&$-$0.147& B0.5 Ia &$-591\pm4$&0.05&$0.03\pm0.05$\\
&$\hspace{-0.15 cm}\phantom{\star w}$4032.28$\tablenotemark{t}$$\star$&J004032.28+404506.7&17.916&$-$0.001& B8 Ia & $-578\pm4$&0.03&$-0.03\pm0.04$\pp\\
&$\phantom{\star}$~4032.88~$\star$&J004032.88+404509.9&17.952&$-$0.103& B2.5e & $-590\pm8$&0.04&$-0.02\pm0.04$\pp\\
&4032.92&J004032.92+404257.7&18.193&$-$0.092& B0.5 Ia &$-594\pm2$&0.11&$0.05\pm0.05$\\
&4034.61&J004034.61+404326.1&18.669&\pp0.150& B1 Ia & $-561\pm2$&0.34&$0.28\pm0.04$\\
&4034.68&J004034.68+404509.9&18.959&\pp0.037& A0 Ia & $-582\pm3$ &0.03&$-0.03\pm0.04$\pp\\
&4034.80&J004034.80+404533.5&18.378&\pp0.254& B8 Ia &$-584\pm2$&0.28&$0.22\pm0.05$\\
&4037.92&J004037.92+404333.3&18.657&\pp0.056& B1.5 Ia & $-571\pm3$&0.23&$0.17\pm0.04$\\
\\
3 (OB8)&4405.89$\tablenotemark{f}$&J004405.89+413016.3&17.309&\pp0.847& K0 & \nodata &0.03&\nodata\\
&4406.87$\tablenotemark{t}$&J004406.87+413152.1&19.019&$-$0.116& B1.5 Ia& \phantom{1}$-59\pm4$&0.05&$-0.01\pm0.04$\pp\\
&4408.04&J004408.04+413258.7&18.708&$-$0.017& B3 Ia& $-105\pm3$&0.11&$0.05\pm0.04$\\
&4408.36&J004408.36+413210.2&19.339&$-$0.098& B0.7 Ia & $-104\pm4$&0.10&$0.04\pm0.05$\\
&$\phantom{\star}$~4408.79~$\star$&J004408.79+413142.0&19.602&$-$0.050& O8-9 II &$-145\pm1$&0.24&$0.18\pm0.06$\\
&4408.94&J004408.94+413201.2&18.356&$-$0.088& B1.5 Ia& $-127\pm5$&0.08&$0.02\pm0.04$\\
&4409.52&J004409.52+413358.9&19.302&$-$0.068& B2 Ib &\phantom{1}$-97\pm4$&0.12&$0.06\pm0.05$\\
&4409.71$\tablenotemark{f}$&J004409.71+413247.2&16.834&\pp0.547& F8 & \nodata &0.00&\nodata \\
&4412.17&J004412.17+413324.2&17.330&\pp0.345& B2.5 Ia &$-110\pm5$&0.49&$0.43\pm0.04$\\
&$\phantom{\star}$~4417.80~$\star$&J004417.80+413408.0&19.397&$-$0.055& O9.5 Ib &\phantom{1}$-94\pm11$&0.19&$0.13\pm0.05$\\
\enddata
\tablenotetext{$\star$}{Possibly blended with other objects.}
\tablenotetext{t}{Stars used as telluric standards.}
\tablenotetext{f}{Foreground stars.}
\tablecomments{Local Group Galaxies Survey (LGGS) designations and photometry are from \citet{massey06}, abbreviated names used in this paper (Abbr.) are also given. Total line-of-sight reddenings (\ebv\ in magnitudes) and radial velocities ($v_{LSR}$ in \kms) are given. M\,31 reddenings (\ebvm) assume a foreground reddening of 0.06~mag as discussed in Section \ref{sec:fg}. Stars suffering from possible spectral and/or photometric contamination due to blending with nearby objects (as determined from the LGGS $V$-band images; \citealt{massey06}), are labelled with a star. The spectrum of 3945.82 was kindly classified by Prof.~P. Crowther (private communication). Two stars in Field 2 have negative \ebvm\ values outside of the error bars, which may be indicative of a problem with their photometry, perhaps due to unseen multiplicity or variations in the assumed foreground reddening.}
\end{deluxetable*}

Observations were carried out between August 2007 and January 2008 as part of queue-program GN-2007B-Q-116. Spectra were obtained in separate blue and red exposures, using the B1200 and R831 gratings respectively, to obtain wavelength coverage from $\sim$\,4000-7000~\AA. Using a slit-width of $1\arcsec$, the mean FWHM of the calibration arc lines was 1.5~\AA\ across the blue spectral range and 2.3~\AA\ in the red. The resulting resolving power is variable as a function of wavelength and ranges from approximately 2500 in the blue to 3500 in the red. Seven or eight $\sim$\,30-minute exposures were obtained for each field in each wavelength range. Total integration times were as follows: Field 1: 3.9~hr (blue), 3.5~hr (red); Field 2: 4~hr (blue), 4.3~hr (red); Field 3: 4~hr (blue), 3~hr (red).

The spectra were reduced using the IRAF \texttt{gemini.gmos} package\footnote{IRAF is distributed by the National Optical Astronomy Observatories, which are operated by the Association of Universities for Research in Astronomy, Inc., under cooperative agreement with the National Science Foundation.}. Images were first cleaned of cosmic rays then flat-fielded. Individual object images were wavelength-calibrated and distortion-corrected using arc lamp exposures, then background-subtracted before extraction of the one-dimensional spectra. The wavelength calibration accuracy was confirmed by measuring the peak wavelengths of the Na~D sky emission lines. Individual reduced spectra were velocity-corrected to the local standard of rest (LSR) frame before co-addition. The signal-to-noise per pixel (S/N) of the reduced spectra is typically $\sim$\,100 near the centres of the blue and red spectral ranges.

\section{Results}

\subsection{Stellar classifications}

\begin{figure*}
\centering
\epsfig{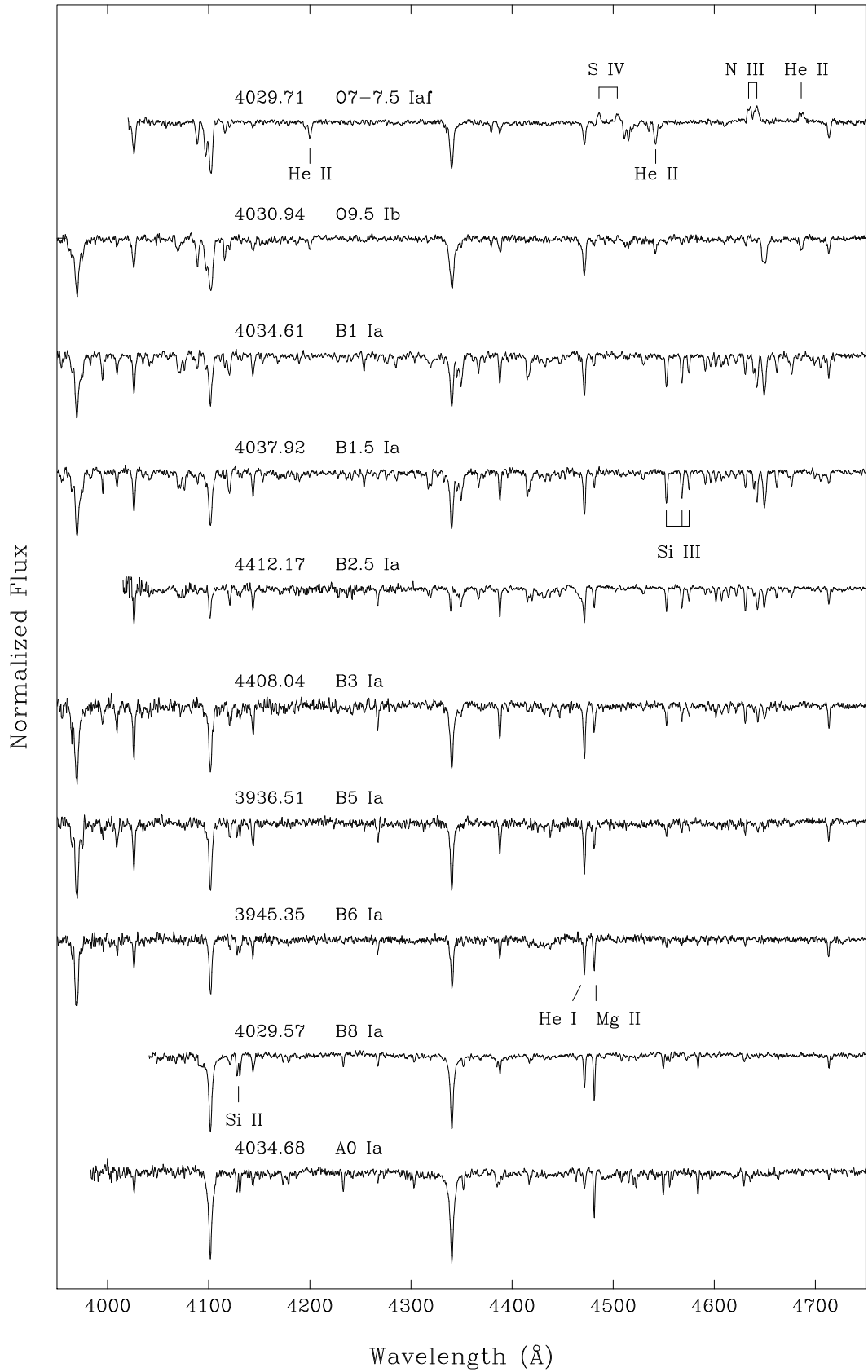}
\caption{Montage of observed M\,31 supergiants over the range of spectral types O7~Iaf to A0~Ia. Note the trends in the
primary diagnostic lines: He\,{\sc i} \lam4471, He\,{\sc ii} \lam\lam4200, 4541, 4686, Mg\,{\sc ii} \lam4481, Si\,{\sc ii}
\lam\lam4128, 4130 \& Si\,{\sc iii} \lam\lam4552, 4560, 4574. Also labelled in the spectrum of 4029.71 are the
Si\,{\sc iv}, N\,{\sc iii} and He\,{\sc ii} emission lines.}
\label{fig:mont}
\end{figure*}

Previous analyses of luminous B-type supergiants in M\,31 found abundances comparable to those in the solar neighbourhood \citep[e.g.][]{trundle02}. The GMOS spectra were therefore classified with respect to Galactic standards \citep{wf90,ldf92}. Spectral types and stellar radial velocities are presented in Table \ref{tab:stars}; those stars with previous spectral classifications are listed for comparison with our new classifications in Table \ref{tab:previous}. Figure~\ref{fig:mont} shows a selection of the blue GMOS spectra, illustrating the spectral sequence of the B-type supergiants.

\begin{deluxetable}{llll}
\tablewidth{0pt}
\tablecaption{New classifications compared to previous work \label{tab:previous}}
\tablehead{
Star & Alias & Sp. Type & Previous
}
\startdata
3945.82 & OB 135-1& WN5 $+$ O7 & WN $\tablenotemark{2}$\\
4029.71 & OB78-231& O7-7.5 Iaf &O8.5 I(f) $\tablenotemark{1}$\\
4030.28 & OB78-277& B1 Ia & B1.5 Ia $\tablenotemark{3}$ \\
4032.88 & OB78-478 & B2.5e & B1.5 Ia $\tablenotemark{4}$ \\
4032.92 & OB78-485 & B0.5 Ia & B0 I $\tablenotemark{5}$\\
4037.92 & \nodata& B1.5 Ia & B0 I $\tablenotemark{6}$\\
4406.87 & OB8-7 & B1.5 Ia & B1 I $\tablenotemark{1}$\\
4408.36 & OB8-25 & B0.7 Ia & B1 I $\tablenotemark{1}$\\
4408.94 & OB8-34 & B1.5 Ia & B1 Ia $\tablenotemark{1}$\\
4409.52 & OB10-43 & B2 Ib & B8 Ia $\tablenotemark{1}$\\
4412.17 & \nodata& B2.5 Ia & B5: I $+$ H{\sc ii} $\tablenotemark{6}$\\
4417.80 & OB9-176 & O9.5 Ib & Early B $\tablenotemark{1}$\\
\enddata
\tablerefs{(1) \citet{massey95}; (2) \citet{mac86}; (3) \citet{bianchi94}; (4) \citet{trundle02}; (5) \citet{hum90}; (6) \citet{massey06}.}
\end{deluxetable}

Reported stellar radial velocities are the mean measurements of the line-centres of selected H, He and metal absorption lines.  The velocity for 4029.71 is a notable outlier of the Field 2 results and the spectrum shows some evidence for a potential second component (e.g. \ion{He}{ii} \lam4200, although in the opposite velocity sense to what might be expected) -- it seems likely this is a composite spectrum of more than one star.

Reddenings (\ebv) were calculated using the intrinsic colours of \citet{wegner94}, supplemented by \citet{johnson66} for later spectral types. The foreground reddening is discussed in Section \ref{sec:fg}.

Observational details of the 34 program stars are summarised in Table \ref{tab:stars}. We adopt abbreviated target names based on the minutes and seconds of right ascension in the LGGS catalogue.  Thus, stars belonging to each of the three fields are distinguished by the first two digits of their abbreviated names (39 for Field 1, 40 for Field 2, and 44 for Field 3).

\subsection{DIB sightlines} \label{sec:selec}

Of the stars observed, two are late-type foreground stars (4405.89 and 4409.71) and one (3945.82) is a Wolf-Rayet in M\,31, and therefore unsuitable for our DIB survey. To investigate the DIBs in M\,31, only the sightlines passing through a significant quantity of ISM were selected for further analysis. Each spectrum was examined for the presence of DIBs, by reference to the survey of \citet{hob08}. No DIBs were detected towards stars with a foreground-corrected reddening of \ebvm~$<0.05$~mag. This is as expected given the spectral noise, and the assumption that M\,31 DIB strengths are similar to those typically observed in the Milky Way: For the typical signal-to-noise (S/N) of the spectra near 5780 and 6283 \AA\ (100 and 80, respectively), using the approximate minimum DIB $W$ detection threshold equation of \citet{hob08} ($W_{max}=1.064\times$FWHM/(S/N)), yields respective lower detection limits of 22 and 63 m\AA\ for these DIBs.  Scaled to a reddening of \ebv~=~$0.04$, the mean Galactic \lam\lam5780 and 6283 DIB $W/E_{B-V}$ values, are 21 and 47 m\AA\ respectively (using data from \citealt{herbig93}, \citealt{thorb03}, \citealt{megier05} and \citealt{cord06}), which is consistent with their non-detection in our spectra at M\,31 reddenings of 0.04 and less. The sample of 14 stars with \ebvm~$\geq0.05$ is listed in Table \ref{tab:nai}, with the addition of three targets (3956.94, 4032.28 and 4406.87), which were used as telluric standards for Fields 1, 2 and 3, respectively.

\subsection{Sodium D lines: column densities \& radial velocities}\label{sec:nai}

\begin{figure}
\centering
\epsfig{file=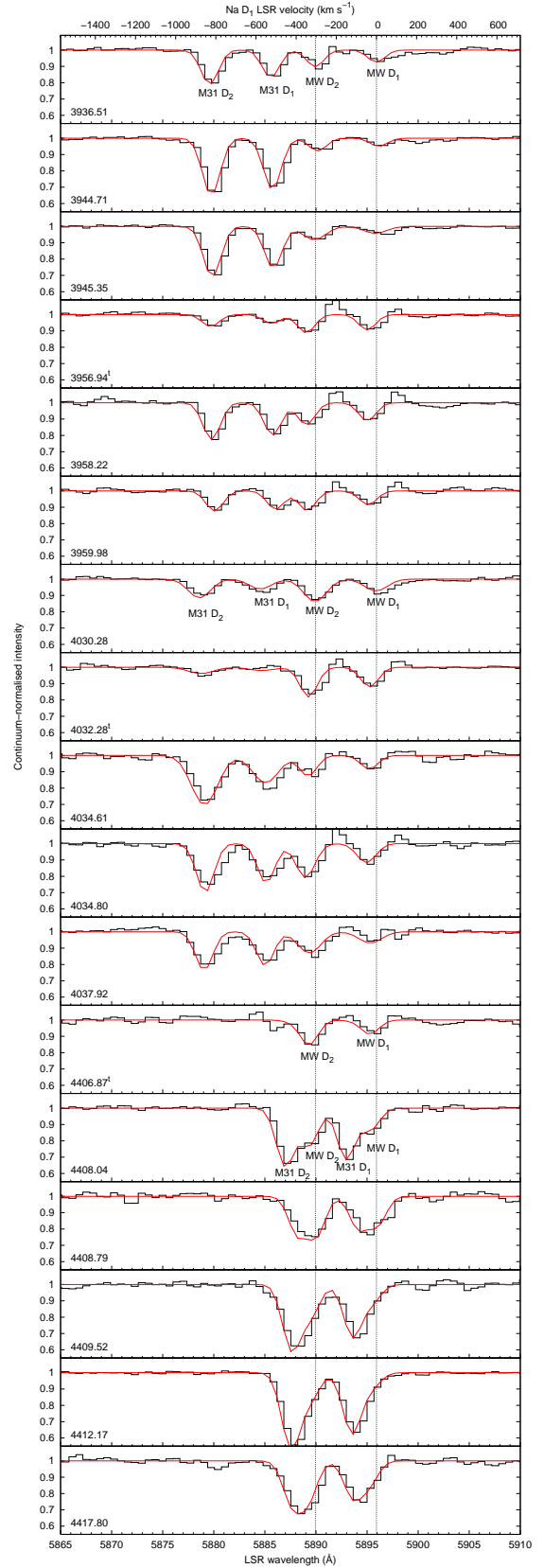, width=0.9\columnwidth}
\caption{Observed continuum-normalised \ion{Na}{i} spectra (histograms) with best-fitting interstellar cloud models overlaid. The velocity scale at the top is relative to the Na D$_1$ rest wavelength. Stellar He \lam5876 lines have been removed from the observed spectra using Gaussian fits. Milky Way (MW) and M\,31 absorption components are marked. Dotted lines show the Na D rest wavelengths. $^{\rm t}$Denotes stars used as telluric standards.}
\label{fig:nai}
\end{figure}

\begin{deluxetable*}{llllllll}
\tablewidth{0pt}
\tablecaption{Best-fit interstellar \ion{Na}{i} cloud model parameters for the Galactic (MW) and M\,31 ISM \label{tab:nai}}
\tablehead{
&\multicolumn{3}{c}{MW}&&\multicolumn{3}{c}{M\,31}\\
\cline{2-4}\cline{6-8}
Sightline&\pp\pp$v$&\pp\pp$b$&\pp\pp$N$&&\pp\pp$v$&\pp\pp$b$&\pp\pp$N$
}
\startdata
3936.51&$-0.4^{+4.1}_{-5.6}$&$4.8^{+3.7}_{-1.7}$&$12.43^{+0.28}_{-0.22}$&&$-526.0^{+2.4}_{-3.0}$&$7.9^{+1.2}_{-1.1}$&$12.90^{+0.14}_{-0.13}$\\[3pt]
3944.71&\pp$11.5^{+5.8}_{-6.2}$&$10.4^{+37.9}_{-7.5}$&$12.14^{+0.18}_{-0.15}$&&$-518.5^{+1.0}_{-1.2}$&$10.5^{+1.6}_{-1.5}$&$13.57^{+0.38}_{-0.29}$\\[3pt]
3945.35&$-4.55^{+3.9}_{-6.9}$&$49.4^{+11.9}_{-13.9}$&$12.09^{+0.03}_{-0.04}$&&$-514.0^{+1.1}_{-1.1}$&$11.1^{+1.0}_{-0.9}$&$13.16^{+0.11}_{-0.10}$\\[3pt]
3956.94$\tablenotemark{t}$&$-46^{+4.6}_{-5.5}$&$3.8^{+1.4}_{-0.9}$&$12.77^{+0.45}_{-0.27}$&&$-521.4^{+7.8}_{-8.1}$&$3.7^{+3.1}_{-1.9}$&$12.36^{+0.56}_{-0.33}$\\[3pt]
3958.22&$-45.6^{+7.8}_{-7.8}$&$5.2^{+2.4}_{-2.1}$&$12.79^{+0.69}_{-0.26}$&&$-516.6^{+4.1}_{-4.6}$&$7.0^{+2.2}_{-1.4}$&$13.29^{+0.45}_{-0.41}$\\[3pt]
3959.98&$-41.2^{+4.8}_{-6.5}$&$5.9^{+3.9}_{-2.0}$&$12.47^{+0.27}_{-0.18}$&&$-502.5^{+3.7}_{-4.7}$&$3.0^{+0.4}_{-0.4}$&$13.56^{+0.63}_{-0.42}$\\[3pt]
\\
4030.28&$-2.0^{+3.0}_{-3.8}$&$51.3^{+7.3}_{-7.0}$&$12.33^{+0.03}_{-0.02}$&&$-578.5^{+3.8}_{-5.1}$&$46.8^{+7.3}_{-10.6}$&$12.24^{+0.03}_{-0.03}$\\[3pt]
4032.28$\tablenotemark{t}$&$-34.1^{+3.0}_{-3.2}$&$10.2^{+5.3}_{-2.8}$&$12.58^{+0.09}_{-0.11}$&&$-571.7^{+22.5}_{-16.3}$&$58.2^{+29.6}_{-40.5}$&$11.79^{+0.12}_{-0.10}$\\[3pt]
4034.61&$-35.7^{+5.4}_{-6.3}$&$8.5^{+21.1}_{-6.3}$&$12.41^{+0.24}_{-0.19}$&&$-552.2^{+2.8}_{-2.8}$&$62.2^{+6.9}_{-6.6}$&$12.78^{+0.02}_{-0.02}$\\[3pt]
4034.80&$-49.6^{+3.5}_{-3.7}$&$26.0^{+8.9}_{-10.2}$&$12.50^{+0.04}_{-0.03}$&&$-548.8^{+2.8}_{-2.6}$&$11.6^{+42.6}_{-2.3}$&$13.08^{+0.22}_{-0.35}$\\[3pt]
4037.92&$-29.8^{+4.9}_{-7.4}$&$49.7^{+13.8}_{-15.2}$&$12.31^{+0.04}_{-0.04}$&&$-554.2^{+3.1}_{-3.6}$&$7.4^{+2.0}_{-1.6}$&$13.26^{+0.51}_{-0.35}$\\[3pt]
\\
4406.87$\tablenotemark{t}$&$-34.0^{+3.2}_{-3.5}$&$16.9^{+7.5}_{-6.0}$&$12.34^{+0.04}_{-0.03}$&&$-121.0\tablenotemark{f}$&$12.0\tablenotemark{f}$&$<11.4$\\[3pt]
4408.04&$-32.0^{+5.2}_{-4.2}$&$22.7^{+10.2}_{-11.0}$&$12.55^{+0.12}_{-0.05}$&&$-151.3^{+2.0}_{-1.9}$&$10.9^{+1.9}_{-2.7}$&$13.48^{+0.62}_{-0.24}$\\[3pt]
4408.79&\pp$4.4^{+10.2}_{-11.6}$&$10.5^{+15.5}_{-5.6}$&$12.81^{+0.49}_{-0.24}$&&$-90.5^{+10.7}_{-13.2}$&$7.7^{+3.2}_{-3.5}$&$13.10^{+0.95}_{-0.36}$\\[3pt]
4409.52&$-23.2\tablenotemark{f}$&$19.9\tablenotemark{f}$&$12.43^{+0.05}_{-0.06}$&&$-119.2^{+2.5}_{-2.2}$&$16.2^{+2.6}_{-3.0}$&$13.16^{+0.18}_{-0.10}$\\[3pt]
4412.17&$-23.2\tablenotemark{f}$&$19.9\tablenotemark{f}$&$12.28^{+0.20}_{-0.24}$&&$-122.9^{+0.7}_{-0.7}$&$18.5^{+0.9}_{-1.1}$&$13.25^{+0.05}_{-0.03}$\\[3pt]
4417.80&$-40.4^{+6.0}_{-6.4}$&$23.4^{+9.4}_{-6.0}$&$12.60^{+0.12}_{-0.05}$&&$-121.1^{+5.3}_{-8.3}$&$7.5^{+1.6}_{-2.4}$&$13.22^{+0.47}_{-0.27}$\\[4pt]
\enddata
\tablenotetext{f}{Values held fixed during the fitting.}
\tablenotetext{t}{Telluric standard stars.}
\tablecomments{Radial LSR velocities ($v$) and Doppler $b$ parameters are in \kms. Column densities ($N$) are in $\log$~cm$^{-2}$. Statistical Monte Carlo errors are given. The MW model parameters may be uncertain due to contamination of the Na D lines near $v=0$ by sky subtraction residuals. M\,31 column densities should generally be considered as lower limits due to the likely presence of unresolved saturated structure in the Na~D lines.}
\end{deluxetable*}

At the resolution of the GMOS spectra (100~\kms\ in this wavelength region), the interstellar Na D absorption lines in M\,31 and the MW can be modelled using single Gaussian interstellar cloud components. Both lines of the Na D doublet were fitted simultaneously using the \textsc{vapid} routine \citep{howarth02}, to produce models of the Galactic and M\,31 \ion{Na}{i} distributions in each sightline. Fitting both lines of the sodium doublet simultaneously helps to constrain the Doppler $b$ parameter in data at this relatively low spectral resolution. The resulting interstellar \ion{Na}{i} cloud radial velocities, Doppler $b$ parameters and column densities are summarised in Table \ref{tab:nai}. Monte Carlo error estimates were generated by refitting each cloud model to 100 different replicated spectra, each with random Poisson noise added, and taking the $\pm68$th percentiles of the resulting (free) parameter ranges.

The Doppler shift of Field 3 is comparable to the instrumental resolving power, and as a result, the MW and M\,31 absorption components are blended which hinders the accurate derivation of the individual component parameters. To obtain physically realistic results for 4409.52 and 4412.17, the velocity and Doppler $b$ parameter of the MW component had to be held fixed during the fitting (at $v=-23.3$~\kms\ and $b=19.9$ \kms, which are the averages for the MW components towards the other stars in Table \ref{tab:nai}).  Towards the telluric standard 4406.87, there was no detectable M\,31 component of \ion{Na}{i}; the only absorption detected is at a mean radial velocity of $-32.0^{+5.2}_{-4.2}$~\kms\ which is unlikely to be in M\,31 and is consistent (in velocity and column density) with the other MW foreground components observed towards the stars in Fields 1 and 2. An upper limit on the M\,31 \ion{Na}{i} column density of \cold{2.8}{11} was derived for this sightline assuming the radial velocity and Doppler width of the gas to be the average of those values measured towards the other stars in Field 3.  It must be noted, however, that all of the Field 3 \ion{Na}{i} fit parameters are subject to additional uncertainty due to the likely contamination of the spectra by sky-subtraction residuals.  Strong telluric \ion{Na}{i} emission features dominate these GMOS spectra at velocities $\pm100$~\kms\ with respect to the D-line centres, and their imperfect subtraction during data reduction thus leads to potential uncertainties in the spectra over this velocity range.

The observed sightlines pass through the halo of M\,31 and are expected to intercept multiple interstellar clouds with component structure that is unresolved in our spectra. Thus, the derived (single-component) interstellar cloud models provide no information about the detailed kinematics of the M\,31 ISM except for the overall velocity width and peak radial velocity of the \ion{Na}{i} gas distributions. The column densities derived from these broad, single-component fits should generally be considered as lower limits due to the likely presence of unresolved saturated structure in the Na~D lines.

Since sodium lines can be present in the background stellar spectra, we investigated the possible stellar Na~D contributions to our observed spectra. Theoretical spectra for a range of stellar types spanning those observed were calculated using \textsc{synspec} \citep{hub00} and \citet{kur93} model atmospheres with solar metallicity. Stellar contributions to the Na D line equivalent widths were found to be negligible compared with the interstellar component for all the stars listed in Table \ref{tab:nai}; only in cooler stars (of type A and later) does the stellar contribution become important.

\begin{figure}
\centering
\epsfig{file=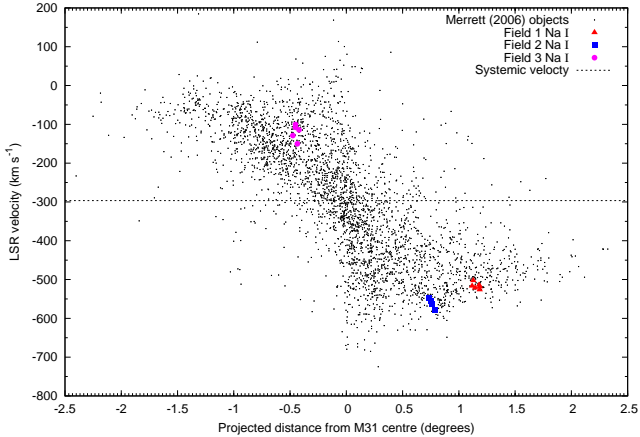, angle=270, width=\columnwidth}
\caption{Kinematic (position-velocity) map of M\,31 along the galaxy's major axis, showing the emission-line objects (mostly planetary nebulae) from the survey by \citet{merrett06} and our measured interstellar \ion{Na}{i} velocities. Data points for each GMOS field are plotted using a different symbol.}
\label{fig:velocity}
\end{figure}

The measured M\,31 interstellar sodium radial velocities match the velocities of nearby M\,31 stars and planetary nebulae, as can be seen in Fig.~\ref{fig:velocity}. This demonstrates with little ambiguity that these clouds are located in the main disc of M\,31.

\subsection{Foreground reddening and DIBs in the Galactic halo} \label{sec:fg}

The mean foreground \ion{Na}{i} column density towards our surveyed M\,31 stars is $3.2\times10^{12}$~cm$^{-2}$.  \citet{sem93} measured the \ion{Na}{i} column density in lines of sight through the Galactic halo, probing gas that should be similar to the MW foreground ISM in our sightlines.  The correlation between \ion{Na}{i} and \ebv\ permits an estimate of the foreground reddening. Performing a least-squares fit to the \citet{sem93} data yields a linear relationship: ${E_{B-V}}=(N({\rm Na~\textsc{i}})+1.41\times10^{11})/3.79\times10^{13}$, from which we derive a foreground reddening of ${E_{B-V}}=0.088$~mag. However, there is considerable scatter in the \ion{Na}{i} \emph{vs.} \ebv\ data -- the mean RMS error on \ebv\ is 0.053 -- and there is likely to be contamination of our \ion{Na}{i} profiles near $v=0$ due to sky-line subtraction residuals. These factors makes the derivation of foreground reddening from \ion{Na}{i} quite uncertain. From the LAB \ion{H}{i} survey\footnote{\mbox{http://www.astro.uni-bonn.de/$\sim$webrai/english/tools\_labsurvey.php}} \citep{kalberla05,hartmann97}, the spectra obtained at the points nearest the centres of our three fields give foreground \ion{H}{i} column densities of $5.75 \times 10^{20}$~cm$^{-2}$ (Field 1), $6.07\times 10^{20}$~cm$^{-2}$ (Field 2), and $6.68\times 10^{20}$~cm$^{-2}$ (Field 3), assuming optically thin emission. Using Equation 7 of \citet{burst78} (which assumes a constant gas-to-dust ratio) to calculate the foreground reddening, the average value is \ebv~=~0.068~mag. The modified \citeauthor{burst78} relation utilised by the NED database\footnote{The NASA/IPAC Extragalactic Database (NED) is operated by the Jet Propulsion Laboratory, California Institute of Technology, under contract with the National Aeronautics and Space Administration.} results in \ebv~=0.0625~mag, which is in good agreement with the foreground reddening towards M\,31 estimated by \citet{schleg98} from the median Galactic dust emission in the surrounding regions (\ebv~=~0.062~mag). We therefore adopt a MW foreground reddening of $0.06\pm0.02$, which is used in the calculation of the foreground-corrected reddenings (\ebvm) given in Table \ref{tab:stars}. Quoted errors on \ebvm\ include an uncertainty of 0.02 in the foreground reddening, plus the uncertainty on the intrinsic stellar $B-V$ values (equal to the uncertainty of a single spectral subclass).

As shown in Figure \ref{fig:5780}, the \lam5780 DIB is detected in several of the observed sightlines at a wavelength matching the Galactic \lam5780 rest wavelength of 5780.37~\AA\ \citep[see][for example]{galaz00}.  This absorption feature most likely originates in the Galactic halo. The average foreground \lam5780 equivalent width is $40\pm10$~m\AA, as measured from the co-added spectrum of all the stars from Field 1 and 2 listed in Table \ref{tab:stars}, (excluding the Wolf-Rayet and those sightlines in Field 3 with blended MW and M\,31 \ion{Na}{i} cloud components). Using a Gaussian fit, the mean radial velocity of the \lam5780 carrier-gas was found to be $0.5\pm7.3$~\kms. The mean Galactic \lam5780 equivalent width per unit reddening is $[W(5780)/E_{B-V}]_{mean}= 527$~m\AA\,mag$^{-1}$, with a standard deviation of 230~m\AA\,mag$^{-1}$ (see Section \ref{sec:selec} for details of the literature data used; the outliers towards $\rho$ Oph and the Orion nebula have been excluded). Using this ratio, the average Milky Way foreground \lam5780 equivalent width measured towards M\,31 corresponds to a foreground reddening of \ebv~=~$0.08^{+0.08}_{-0.04}$~mag, which is consistent with the value of 0.06 adopted in this study.

\subsection{M\,31 diffuse interstellar bands: equivalent widths and radial velocities}\label{sec:dibs}

\begin{figure}
\centering
\epsfig{file=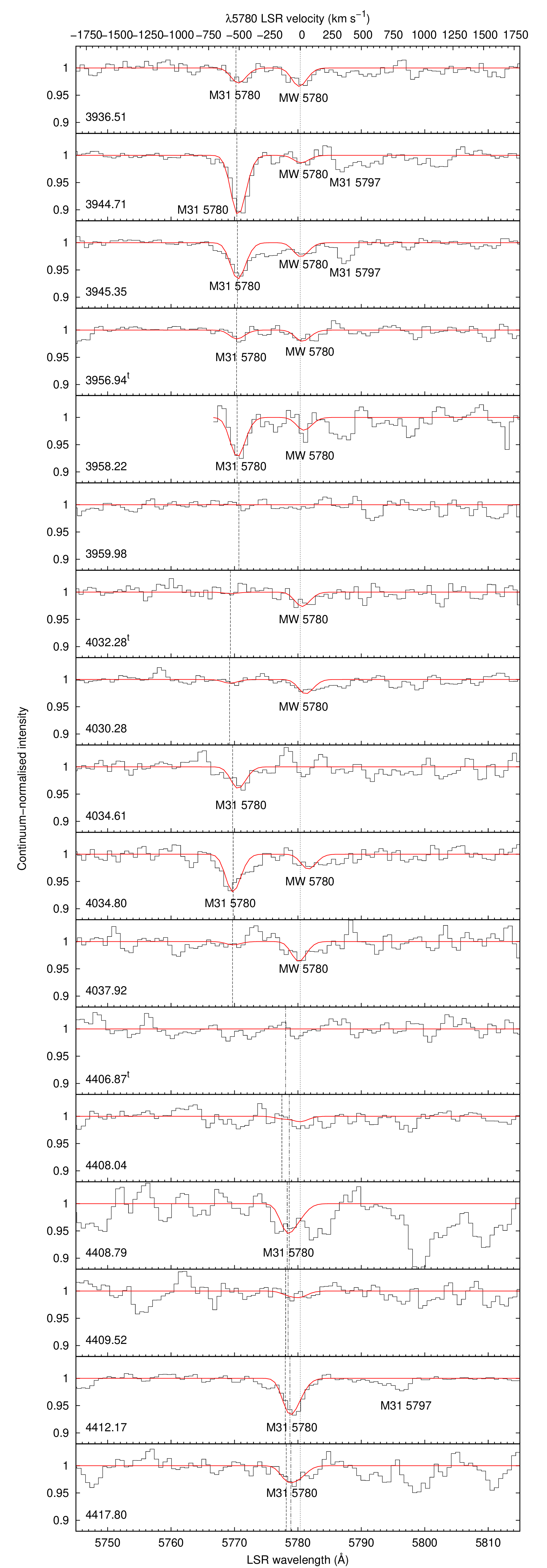, width=0.9\columnwidth}
\caption{Observed continuum-normalised spectra of the \lam5780 region (histograms). \lam5780 components arising in MW and M\,31 have been fitted using the Galactic sightline $\beta^1$ Sco as a template, (fitted profiles overlaid in red). Velocity scale at the top is relative to the Galactic \lam5780 rest wavelength of 5780.37 \AA\ \citep{galaz00} (plotted with a vertical dotted line). The mean M\,31 interstellar \ion{Na}{i} absorption wavelengths are plotted with vertical dashed lines. For the Field 3 stars, the peak \ion{H}{i} velocities are also shown (dot-dashed lines). $^{\rm t}$Denotes telluric standard stars.}
\label{fig:5780}
\end{figure}

\begin{figure}
\centering
\epsfig{file=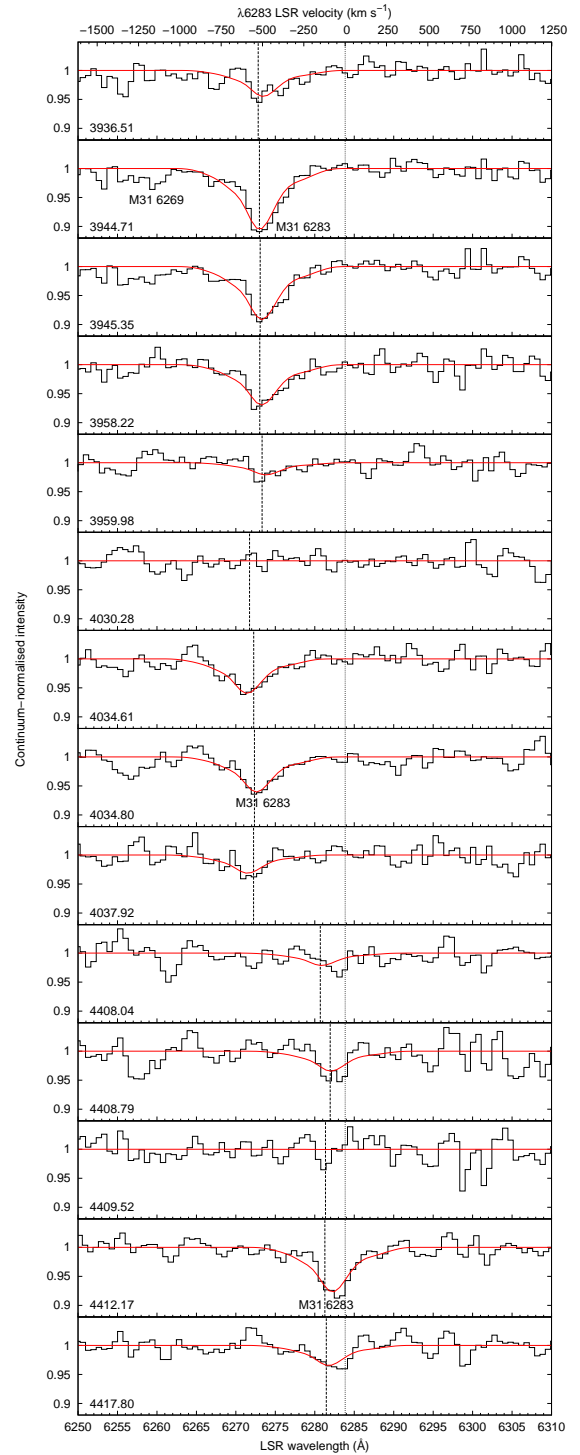, width=0.9\columnwidth}
\caption{Observed continuum-normalised, telluric-corrected spectra of the \lam6283 region (histograms), with fitted DIB profiles overlaid (red curves). Velocity scale at the top is relative to the Galactic \lam6283 rest wavelength of 6283.85 \AA\ \citep{galaz00} (plotted with a vertical dotted line). The mean M\,31 interstellar \ion{Na}{i} absorption wavelengths are plotted with vertical dashed lines.}
\label{fig:6283}
\end{figure}

\tabletypesize{\scriptsize}
\def\arraystretch{1.4}
\begin{deluxetable*}{lccccccccccccccc}
\tablewidth{0pt}
\tabletypesize{\tiny}
\tablecaption{\lam5780 and \lam6283 DIB equivalent widths and velocities, M\,31 \ion{H}{i} 21~cm column densities, gas-to-dust ratios, interstellar NUV radiation field strengths and Spitzer 8/24~$\mu$m flux ratios for observed sightlines \label{tab:5780}}
\tablehead{&\multicolumn{2}{c}{$W(5780)$}&&\multicolumn{2}{c}{$W(6283)$}&&&&&&&\\
\cline{2-3}\cline{5-6}
&Fit&Int.&&Fit&Int.&&$v(5780)$&$v(6283)$&&\n{\ion{H}{i}}&$G$&$I_{NUV}$&8/24~$\mu$m&$W(5780)/E_{B-V}$&$W(6283)/E_{B-V}$\\
Sightline&m\AA&m\AA&&m\AA&m\AA&&\kms&\kms&&$10^{21}$ cm$^{-2}$&\tablenotemark{a}&Draine&flux&\AA\,mag$^{-1}$&\AA\,mag$^{-1}$
}
\startdata
3936.51&$84^{+33}_{-32}$&89 (19)&&$279^{+96}_{-85}$&217 (55)&&$-510.7^{+36.8}_{-36.0}$&$-498.5^{+63.9}_{-49.3}$&&2.8&20&0.27&1.16&$1.2^{+2.8}_{-0.7}$&$4.0^{+8.9}_{-2.3}$\\
3944.71&$318^{+26}_{-24}$&359 (17)&&$671^{+79}_{-71}$&544 (48)&&$-512.4^{+7.5}_{-7.0}$&$-524.8^{+13.6}_{-15}$&&4.5&7.1&0.48&0.63&$1.0^{+0.2}_{-0.2}$&$2.1^{+0.6}_{-0.4}$\\
3945.35&$\tablenotemark{b}202^{+22}_{-24}$&$\tablenotemark{b}$302 (14)&&$587^{+69}_{-74}$&414 (52)&&$-515.4^{+12.2}_{-11.7}$&$-510.6^{+15.6}_{-15.4}$&&4.2&6.4&0.35&0.94&$0.6^{+0.2}_{-0.1}$&$1.8^{+0.5}_{-0.4}$\\
3958.22&$211^{+33}_{-48}$&207 (35)&&$439^{+98}_{-89}$&419 (60)&&$-517.9^{+18.1}_{-16.7}$&$-511.9^{+26.4}_{-25.4}$&&4.0&8.5&0.35&1.78&$0.9^{+0.4}_{-0.3}$&$1.9^{+0.9}_{-0.6}$\\
3959.98&$<29$& $<42$&&$117^{+90}_{-60}$&124 (54)&&\nodata&$-481.3^{+81.6}_{-51.6}$&&2.6&25&0.59&0.71&$<2.2$&$2.2^{+13.8}_{-1.6}$\\[6pt]
4030.28&$<41$& $<28$&&$<42$&$<112$&&\nodata&\nodata&&2.8&16&3.1&0.90&$<0.9$&$<0.9$\\
4034.61&$120^{+39}_{-41}$&165 (25)&&$375^{+105}_{-89}$&405 (65)&&$-514.9^{+33.3}_{-35}$&$-601.3^{+31.0}_{-30.4}$&&2.3&4.0&2.9&1.20&$0.4^{+0.2}_{-0.2}$&$1.3^{+0.7}_{-0.4}$\\
4034.80&$215^{+24}_{-26}$&238 (22)&&$383^{+76}_{-78}$&326 (54)&&$-584.5^{+14.2}_{-13.7}$&$-543.5^{+25.3}_{-27.4}$&&1.3&2.9&1.9&1.15&$1.0^{+0.4}_{-0.3}$&$1.7^{+0.9}_{-0.6}$\\
4037.92&$<39$& $<49$&&$177^{+122}_{-82}$&123 (68)&&\nodata&$-600.1^{+43.4}_{-49.1}$&&3.3&10&0.97&1.38&$<0.3$&$1.1^{+1.3}_{-0.6}$\\[6pt]
4408.04&$<30$& $<40$&&$<243$& $<142$&&\nodata&\nodata&&1.6&15&0.79&1.16&$<2.3$&$<2.7$\\
4408.79&$\tablenotemark{b}146^{+74}_{-48}$& $<88$&&$189^{+169}_{-115}$&171 (104)&&\nodata&$-93.8^{+39.0}_{-47.0}$&&1.2&3.4&0.83&1.05&$0.8^{+1.0}_{-0.4}$&$1.1^{+1.9}_{-0.7}$\\
4409.52&$<38$& $<57$&&$<65$& $<168$&&\nodata&\nodata&&1.4&11&0.59&1.26&$<3.2$&$<5.4$\\
4412.17&$190^{+17}_{-16}$&256 (10)&&$491^{+83}_{-81}$&423 (53)&&$-88.1^{+8.2}_{-8.2}$&$-82.8^{+17.7}_{-22.3}$&&2.0&2.3&1.1&0.90&$0.8^{+1.0}_{-0.4}$&$1.1^{+1.9}_{-0.7}$\\
4417.80&$79^{+55}_{-41}$& $<57$&&$194^{+148}_{-92}$& 221 (76)&&$-99.3^{+57.8}_{-48.6}$&$-107.0^{+42.3}_{-55.6}$&&1.7&6.6&0.48&0.76&$0.6^{+1.2}_{-0.4}$&$1.6^{+3.0}_{-1.0}$\\
\enddata
\tablenotetext{a}{Units of $G$ are $10^{21}$\,cm$^{-2}$\,mag$^{-1}$}
\tablenotetext{b}{DIB equivalent widths probably affected by line blending.}
\tablecomments{Fitted and continuum-integrated DIB equivalent widths are given, labelled Fit and Int., respectively. The errors ($\pm$) on the integrated equivalent widths are $\sigma\Delta\lambda$ and are given in brackets, with upper limits calculated as $2\sigma\Delta\lambda$ . Errors on the gas-to-dust ratios ($G=\frac{1}{2}$ \n{\ion{H}{i}}/\ebvm) are $\pm100\%$ due to probable contamination by background \ion{H}{i} (see Section \ref{sec:HI}).}
\end{deluxetable*}
\tabletypesize{\footnotesize}
\def\arraystretch{1.2}

We searched the spectra of each of the sightlines with foreground-corrected reddening \ebvm~$\geq0.05$ for the DIBs listed in the survey of \citet{hob08}. For the \lam\lam6283 and 6993 DIBs, which are partially obscured by overlapping telluric lines, corrections were performed by division with spectra from lines-of-sight with negligible \ebvm\ selected from the same field (designated with superscript t in Table \ref{tab:stars}). The \lam\lam4428, 5705, 5780, 5797, 6203, 6269, 6283, 6379, 6613, 6660 and 6993 DIBs were detected in at least one sight-line each, with Doppler shifts matching the M\,31 \ion{Na}{i} and \ion{H}{i} velocities (\ion{H}{i} data are presented in Section \ref{sec:HI}). The \lam5849 and \lam6445 DIBs were tentatively detected in the spectrum of 3944.71 only. The \lam\lam4501, 4726, 6196, 6376 and 6532 DIBs were too weak to be detected in any of the spectra, which is consistent with the assumption of MW-like strengths (per unit reddening), given the S/N and low resolving power of our spectra. Severe contamination by telluric absorption prevented analysis of the \lam\lam6886, 6919, 7224 and 7334 DIBs. Equivalent widths were measured by integration across a linear or quadratic fitted continuum. Errors on DIB equivalent widths measured this way were determined by $\sigma\Delta\lambda$, where $\sigma$ is the RMS noise in the continuum and $\Delta\lambda$ is the full-width half-maximum of each DIB. As a result of the relatively low signal-to-noise and low dispersion of the spectra compared with other contemporary Galactic DIB studies \citep[\eg][]{mcc10}, additional uncertainties in the measured DIB equivalent widths arise due to the difficulties in establishing the true full-widths of the DIBs over which to perform the integration. For example, in Figure \ref{fig:5780} the presence of overlapping weak stellar/noise features at the wings of the \lam5780 DIB observed towards 3944.71, 4034.80 and 4412.17, cause depressions in the continua that cannot be distinguished from the DIB by eye, and thus artificially boost the measured equivalent widths. Conversely, the broad, shallow, Lorentzian-like wings of the \lam6283 DIB (shown in Figure \ref{fig:6283}), are lost in the noise, resulting in likely underestimation of the full-width of this feature. Therefore, to provide a second, more accurate measurement of the \lam\lam5780 and 6283 radial velocities and equivalent widths that takes the complex profiles of the DIBs into account, Galactic DIB template spectra were fitted to the observed M\,31 spectra.

\begin{figure}
\centering
\epsfig{file=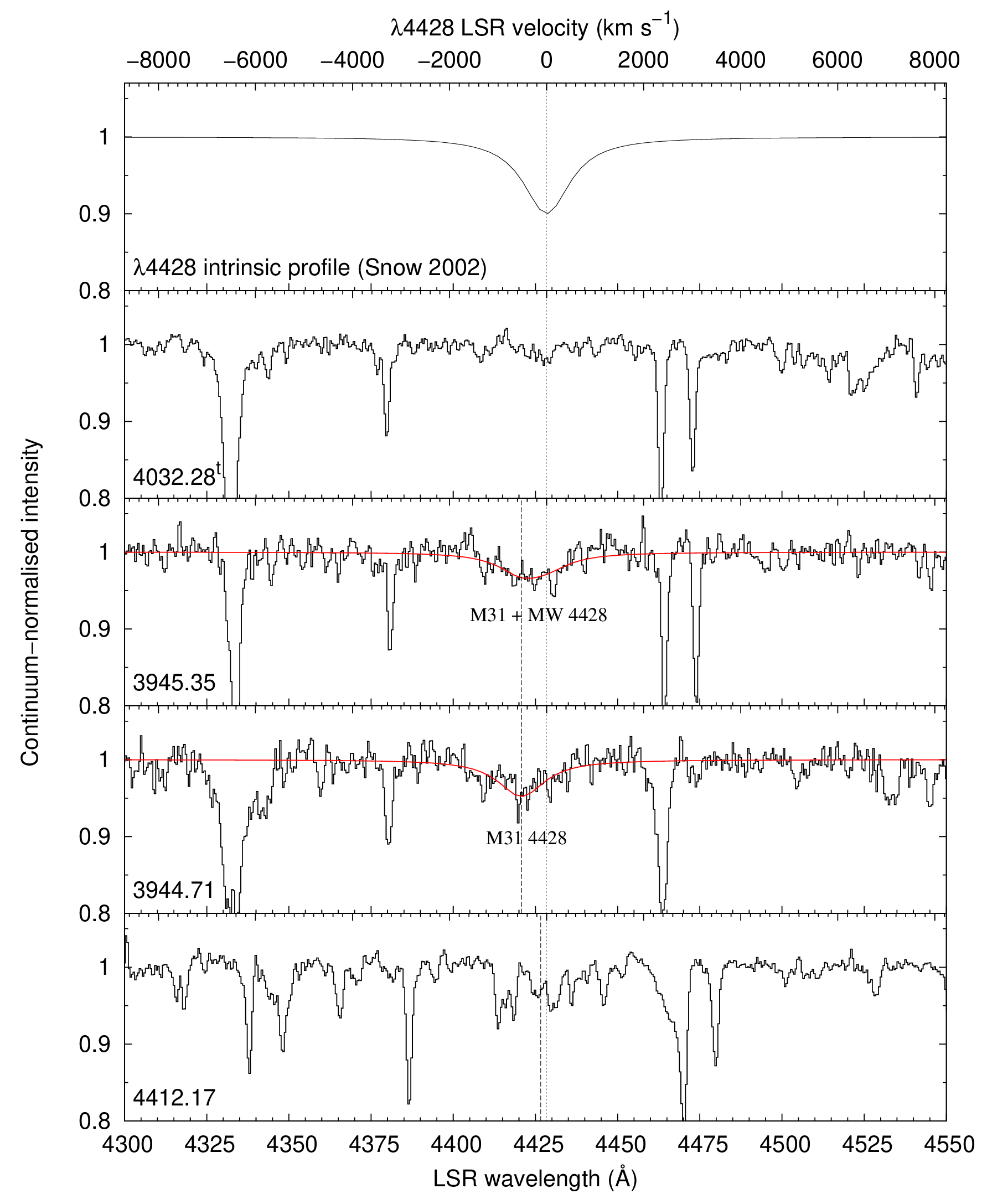, width=0.9\columnwidth}
\caption{Observed continuum-normalised spectra of the \lam4428 region (histograms). Upper panel shows the intrinsic MW \lam4428 profile \citep{snow02}. Velocity scale at the top is relative to the Galactic DIB rest wavelength of 4428.39 \AA\ (dotted line). M\,31 \ion{Na}{i} velocities are plotted with dashed lines.}
\label{fig:4428}
\end{figure}

\begin{figure}
\centering
\epsfig{file=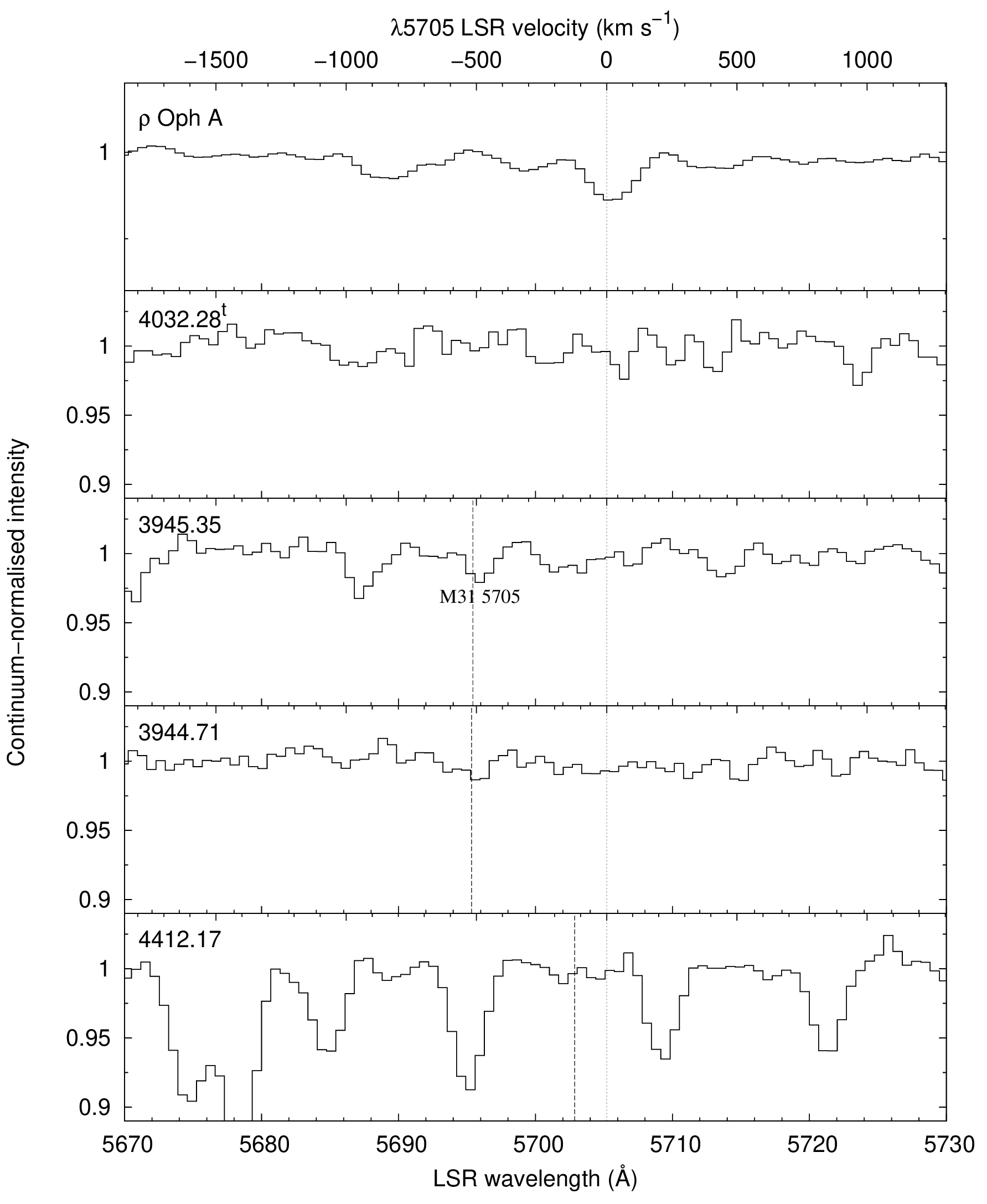, width=0.9\columnwidth}
\caption{Observed continuum-normalised spectra of the \lam5705 region. Upper panel shows MW comparison DIB spectrum towards $\rho$\,Oph A \citep{cord06}. The second panel from the top shows the spectrum of one of the telluric standard stars. Velocity scales at the top of this Figure and Figures \ref{fig:6203}-\ref{fig:6993} are relative to the Galactic DIB rest wavelengths published by \citet{galaz00}. The mean M\,31 interstellar \ion{Na}{i} absorption wavelengths are plotted with vertical dashed lines. The spectrum of 4412.17 is dominated by stellar features at this wavelength.}
\label{fig:5705}
\end{figure}

\begin{figure}
\centering
\epsfig{file=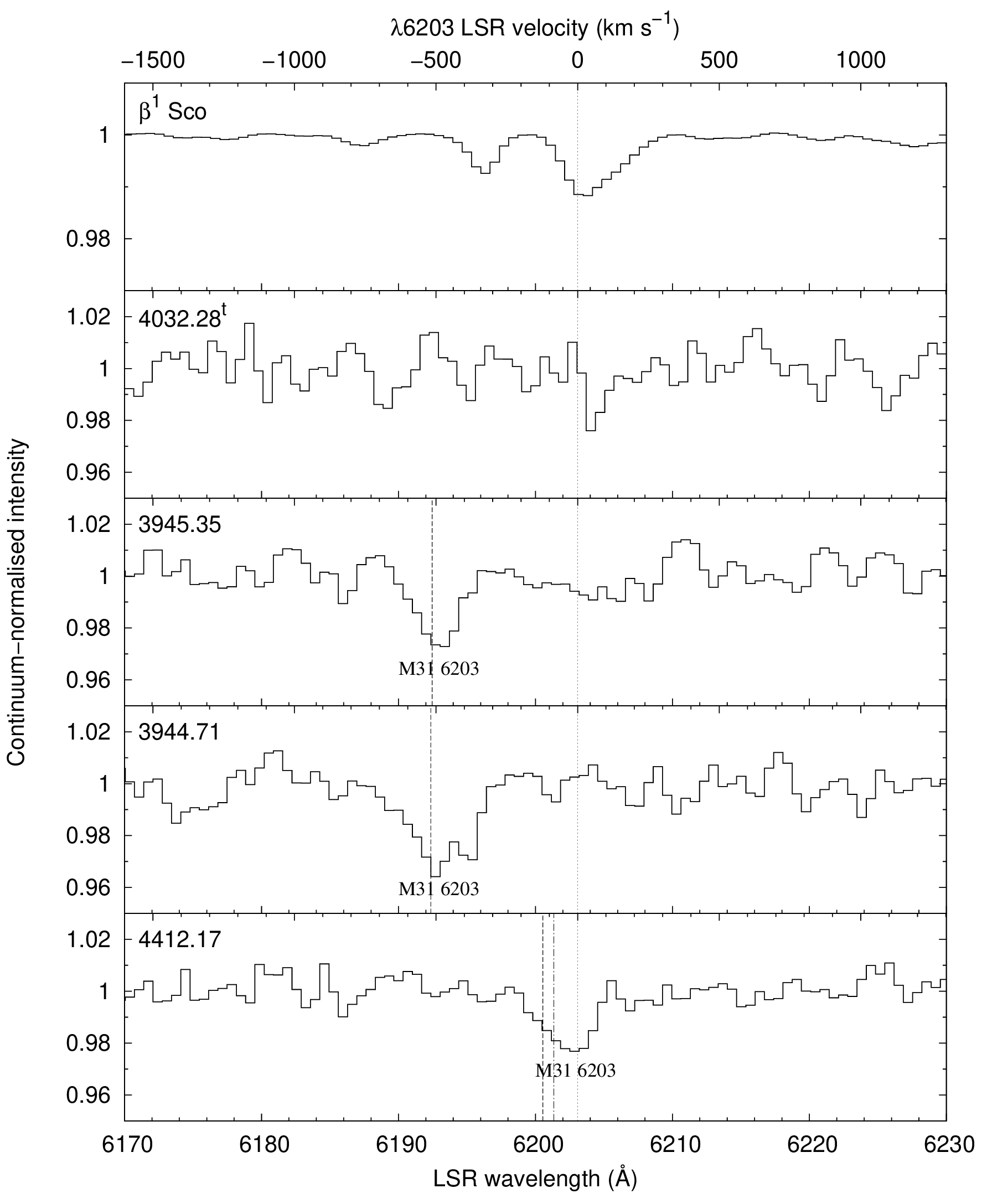, width=0.9\columnwidth}
\caption{Observed continuum-normalised spectra of the \lam6203 region. Upper panel shows MW comparison DIB spectrum towards $\beta^1$\,Sco \citep{cord06}. The DIB rest velocity is shown with a dotted line and the M\,31 \ion{Na}{i} velocities are plotted with dashed lines. For 4412.17, the mean \ion{H}{i} velocity is shown with a dot-dashed line.}
\label{fig:6203}
\end{figure}

\begin{figure}
\centering
\epsfig{file=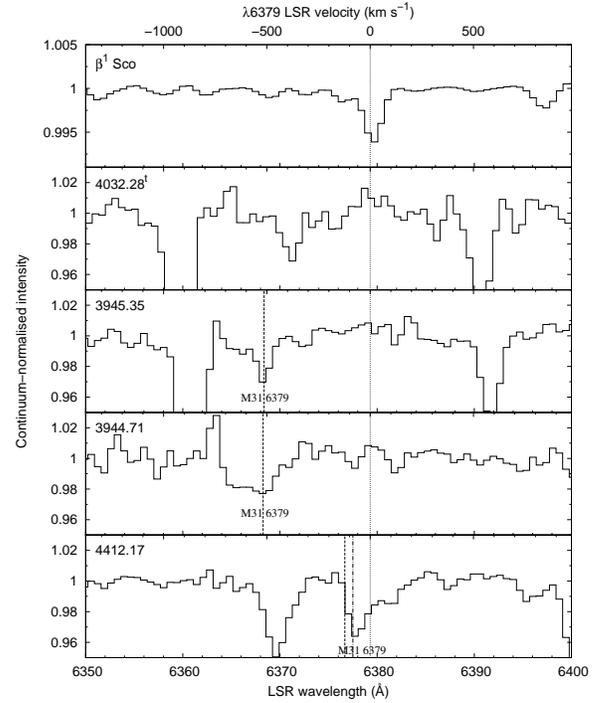, width=0.9\columnwidth}
\caption{Observed continuum-normalised spectra of the \lam6379 region. Upper panel shows MW comparison DIB spectrum towards $\beta^1$\,Sco \citep{cord06}.
The DIB rest velocity is shown (dotted line), and the M\,31 \ion{Na}{i} velocities are plotted with dashed lines. For 4412.17, the mean \ion{H}{i} velocity is shown with a dot-dashed line. The strong absorption features around 6360 and 6390 \AA\ are due to stellar \ion{S}{ii} and \ion{Ne}{i}.}
\label{fig:6379}
\end{figure}

\begin{figure}
\centering
\epsfig{file=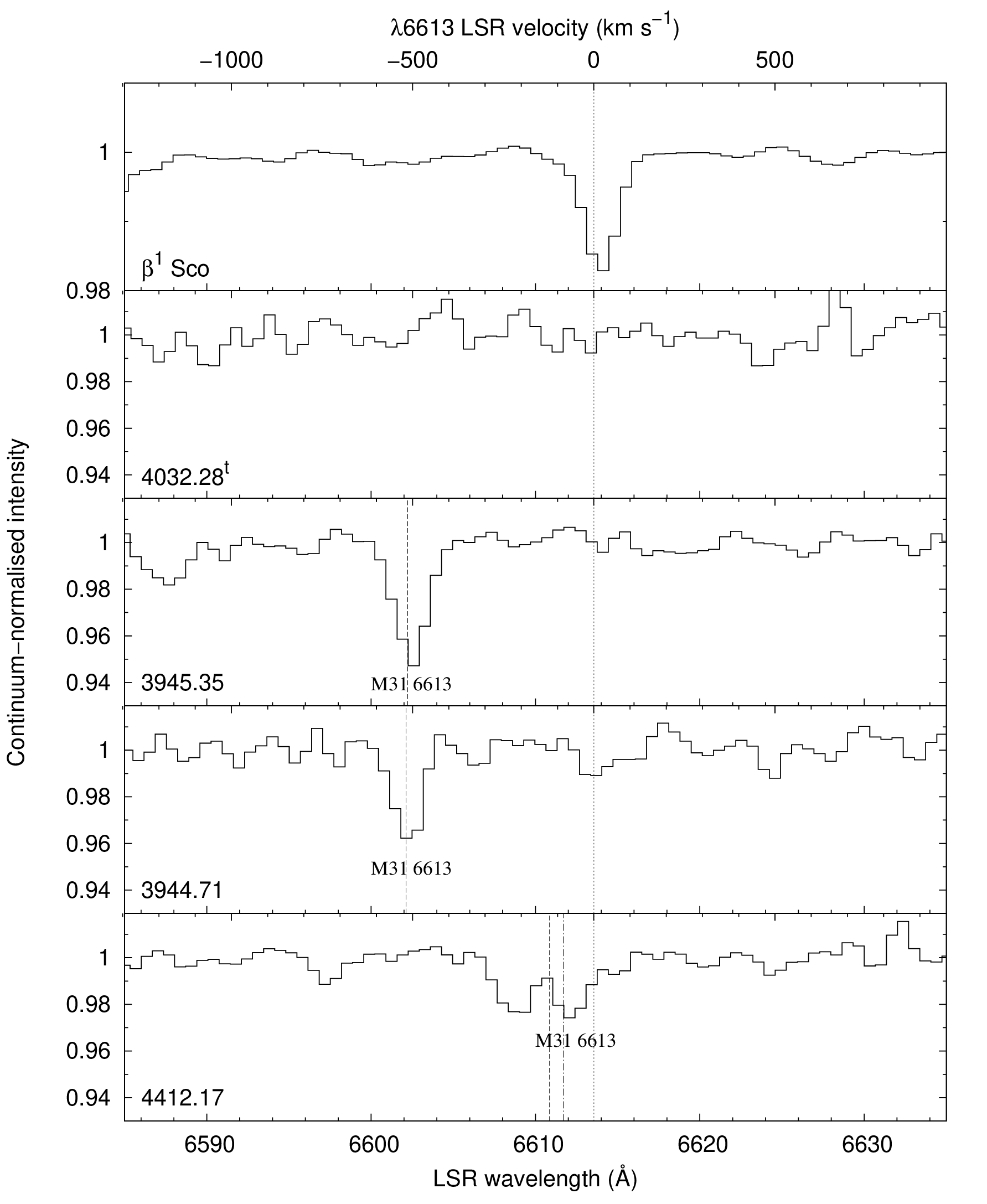, width=0.9\columnwidth}
\caption{Observed continuum-normalised spectra of the \lam6613 region. Upper panel shows MW comparison DIB spectrum towards $\beta^1$\,Sco \citep{cord06}.
The DIB rest velocity is plotted (dotted line), and the M\,31 \ion{Na}{i} velocities are plotted with dashed lines. For 4412.17, the mean \ion{H}{i} velocity is shown with a dot-dashed line.}
\label{fig:6613}
\end{figure}

\begin{figure}
\centering
\epsfig{file=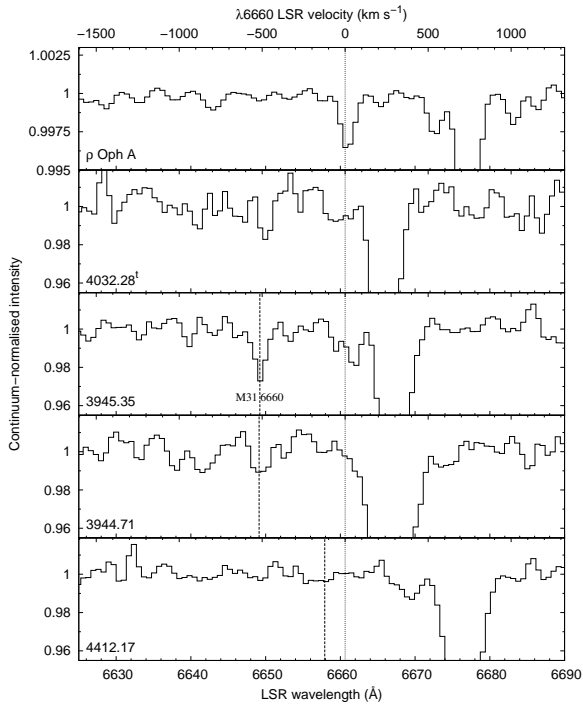, width=0.9\columnwidth}
\caption{Observed continuum-normalised spectra of the \lam6660 region (histograms). Upper panel shows MW comparison DIB spectrum towards $\rho$\,Oph A \citep{cord06}.
The DIB rest velocity and M\,31 \ion{Na}{i} velocities are plotted with dotted lines. The strong stellar line at the red end is due to \ion{He}{i}.}
\label{fig:6660}
\end{figure}

\begin{figure}
\centering
\epsfig{file=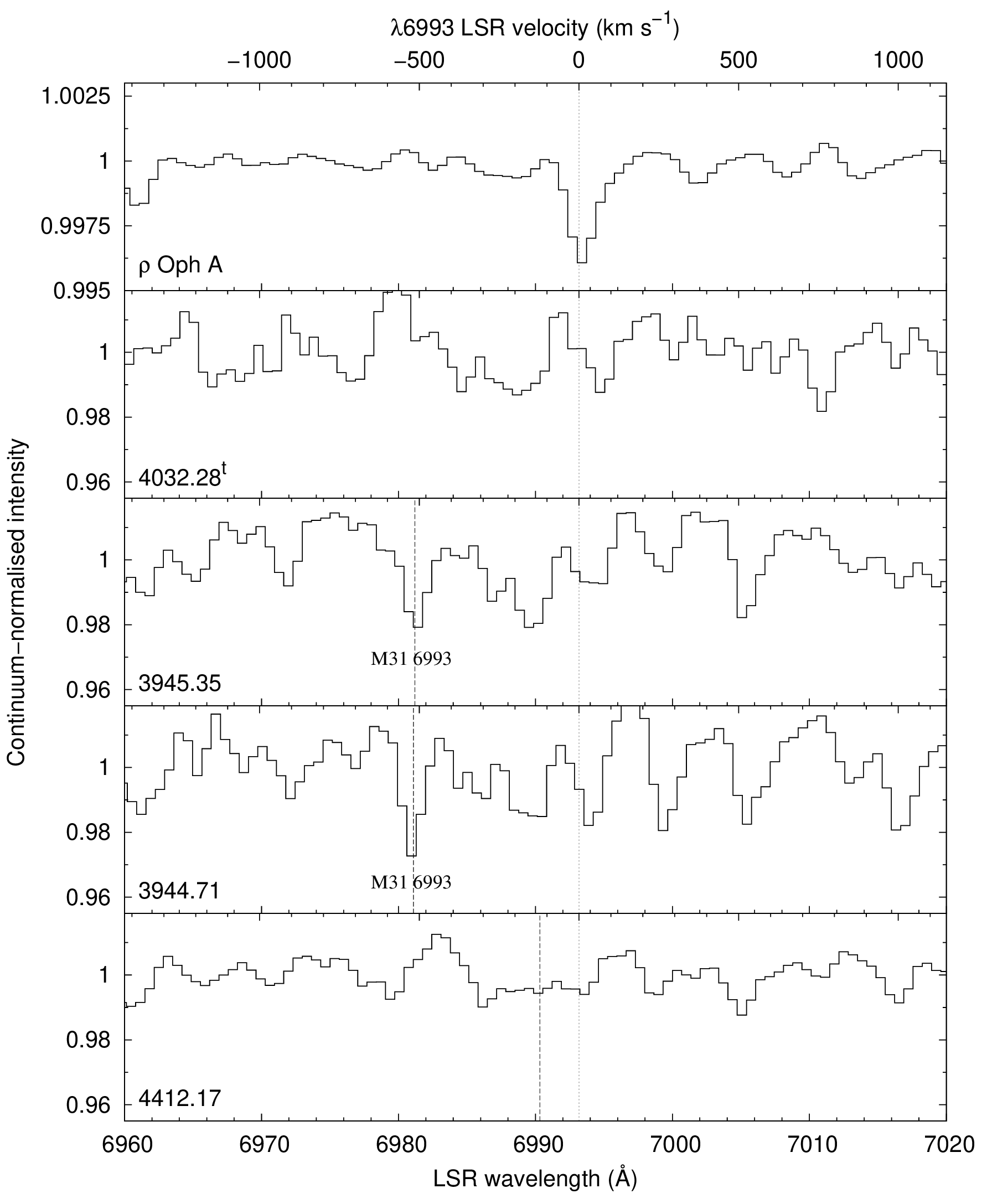, width=0.9\columnwidth}
\caption{Observed continuum-normalised spectra of the \lam6993 region. Upper panel shows MW comparison DIB spectrum towards $\rho$\,Oph A \citep{cord06}. The DIB rest velocity and M\,31 \ion{Na}{i} velocities are plotted with dotted lines.}
\label{fig:6993}
\end{figure}

The Galactic DIB templates were derived from high resolution ($R=58,000$), high S/N ($\sim2000$) spectra of \object[HD144217]{$\beta^1$\,Sco} (see \citealt{cord06}), shifted to the interstellar rest frame. These were convolved with a Gaussian interstellar cloud model, then with the GMOS spectral PSF (assumed Gaussian), and rebinned to the dispersion of the GMOS spectra. The interstellar cloud model was optimised using the same nonlinear least-squares algorithm employed by the \textsc{vapid} code \citep{howarth02}. This technique has the benefit that the statistical uncertainties in the measurements of the equivalent widths and velocities of weak DIBs are minimised, but has the drawback that the intrinsic M\,31 DIB profiles and rest wavelengths are assumed to closely match those of the Milky Way DIB standard star $\beta^1$\,Sco. If this were not the case, unknown systematic errors could occur, but there is no evidence to suggest that M\,31 DIB profiles differ from those of $\beta^1$\,Sco. Profiles of some Galactic DIBs have been shown to exhibit variations in substructure (\citealt{galaz02}). So far, however, all observed variations have been small and are negligible compared to the size of the GMOS resolution element and the level of noise in our spectra. The reader is referred to Section 3.5.2 of \citet{cord06} for a more detailed discussion of the merits of the DIB fitting technique.

The fitted and integrated equivalent widths are shown in Table \ref{tab:dibs}, and match each other to within the error bars. However, for four out of seven cases (for which the DIBs are not affected by obvious blends, denoted by superscript $b$), the fitted values of $W(5780)$ are systematically less than the integrated values. For W(6283), the fitted values are generally greater than the integrated values. These differences are as expected given the increased accuracy afforded by using known DIB profiles to better constrain their full-widths, as explained above.

Error estimates for the \lam5780 and \lam6283 equivalent widths and velocities were derived by the same Monte Carlo technique as used for the Na\,{\sc i} spectral modelling; 500 different replicated spectra were produced for each DIB with (random) Poisson noise added to each, the $\sigma$-value for which was taken to be equal to the RMS noise of the local continuum. Errors due to the uncertainties inherent in the spectral continuum rectification were accounted for by multiplicatively scaling each replicated spectrum by a random factor of between $1-\sigma$ and $1+\sigma$. The DIB model parameters were refitted for each replicated spectrum and individual parameter errors were obtained from the $\pm68$th percentiles of the respective parameter ranges.

As can be seen from the data in Table \ref{tab:dibs}, the error estimates on the DIB equivalent widths derived from Monte Carlo fitting (in the column labeled `Fit.') are typically larger  (by $\sim50$\%) than those derived using $\sigma\Delta\lambda$ (column labeled `Int.'). Although our Monte Carlo errors do not take account of interloping stellar features (these are considered separately; see below), they are expected to be more reliable than the $\sigma\Delta\lambda$ errors, which assume a somewhat arbitrary $\pm1\sigma$ error on the continuum placement, as described by \citet{mcc10}, and do not take into account the non-Gaussianity of the DIB profiles, which is especially important for \lam6283 with its very broad, Lorentzian-like wings.

The spectral regions spanning the \lam5780 and \lam5797 DIBs are shown in Figure \ref{fig:5780} for the fourteen stars in our M\,31 reddened sample (plus three telluric standards). The spectral regions covering \lam6269 and \lam6283 are shown in Figure \ref{fig:6283} (these spectra have been divided by the spectra of the telluric standards from each respective field). M\,31 components of \lam5780 are detected in 10 sightlines and \lam6283 in 11 sightlines. The MW foreground \lam6283 components are effectively removed during telluric division because of their presence in the telluric standard spectra.

A total of 11 different DIBs (\lam\lam4428, 5705, 5780, 5797, 6203, 6269, 6283, 6379, 6613, 6660, and 6993) were detected in M\,31 towards the three most reddened targets 3944.71, 3945.35 and 4412.17 (data presented in Table \ref{tab:dibs} and Figs.~\ref{fig:5780} to~\ref{fig:6993}). Figures~\ref{fig:5705} to~\ref{fig:6993} also show reference MW DIB spectra \citep[from][]{cord06}, as well as telluric standard spectra to illustrate the likely level of contamination by stellar and telluric features.

To account for possible additional uncertainties in the DIB equivalent widths given in Table \ref{tab:dibs} due to the presence of overlapping stellar lines, we have estimated the possible stellar line contributions. The table shows the total equivalent widths of stellar features that might be overlapping the DIBs, calculated from synthetic B0\,Ia and B9\,Ia stellar spectra (at temperatures of 28,000 and 10,000~K, respectively), using the method described in Section \ref{sec:nai}. The line list used in the spectral modeling includes transitions of the following elements and their ions: H, He, C, N, O, Na, Mg, Si, Ca, Cr, Mn and Fe. The calculated stellar line contributions to measured DIB equivalent widths are generally small (except for the very broad \lam4428 DIB which is contaminated by the \ion{O}{ii} blend near 4415 \AA), and in most cases are negligible compared with the errors arising from continuum placement and statistical noise.

\begin{deluxetable}{rllllll}
\tablewidth{0pt}
\tablecaption{DIB equivalent widths for targets with many DIB detections \label{tab:dibs}}
\tablehead{Sightline& \colhead{3944.71} & \colhead{3945.35} & \colhead{4412.17} & \colhead{MW ISM}& \colhead{$\star$} & \colhead{$\star$}\\ \colhead{Sp. type}&\colhead{O9.7\,Ib}&\colhead{B6\,Ia}&\colhead{B2.5\,Ia}&\colhead{avg.}&\colhead{B0\,Ia}&\colhead{B9\,Ia}\\ \ebv&\colhead{0.32}&\colhead{0.33}&\colhead{0.43}&\colhead{0.32}&\colhead{0}&\colhead{0}}
\startdata
$W(4428)$ & 1129 (106) &628 (135) & \tablenotemark{b} & 865&640&370\\
$W(5705)$ & $<31$ &42 (25) & $<19$ & 38&5&8\\
$W(5780)$ & 318 (25) &202 (23)\tablenotemark{b} &190 (17) & 163&2&30\\
$W(5797)$ & 62 (35) &101 (28) &45 (21) & 49&4&9\\
$W(6203)$ & 159 (65) &93 (54) &92 (40) & 54&5&4\\
$W(6269)$ & 60 (23) &\tablenotemark{b} &$<14$ & \dots&0&7\\
$W(6283)$ & 671 (75) &587 (73) &491 (82) & 362&1&10\\
$W(6379)$ & \tablenotemark{b} &25 (18) &63 (13)\tablenotemark{b}& 28&8&1\\
$W(6613)$ &80 (25) &123 (20) &57 (15) & 68&0&5\\
$W(6660)$ &$<25$ &54 (20) &$<15$ & 16&0&1\\
$W(6993)$ &45 (23) &77 (19) &$<14$ & 38&0&2\\
\enddata
\tablenotetext{b}{Uncertain (or unmeasurable) equivalent widths due to line blends.}
\tablecomments{DIB equivalent widths are in m\AA. Average DIB strengths (per unit \ebv) for the Galactic diffuse ISM (scaled to \ebv\ = 0.32) are given in the column labelled MW ISM (data from \citealt{herbig93}, \citealt{thorb03}, \citealt{megier05} and \citealt{cord06}). No large sample of \lam6269 measurements were available. Possible stellar line contributions (in m\AA) to DIB equivalent widths are given in the columns labelled $\star$ for B0\,Ia and B9\,Ia stars, measured from synthetic stellar spectra with temperatures of 28,000 and 10,000~K, respectively.
}
\end{deluxetable}

As a result of the close blending between the MW and M\,31 interstellar gas components due to the low Doppler shift of Field 3 (discussed further in Sections \ref{sec:nai} and \ref{sec:HI}), there may be some ambiguity as to in which galaxy the Field 3 DIBs originate. The main discriminator we use to identify the location of the DIB carriers is their radial velocity compared with the \ion{Na}{i} component structure. The separation of the MW and M\,31 DIB velocities is clear in all cases apart from 4412.17 and 4417.80, for which the DIB velocities appear to be intermediate between the M\,31 and MW \ion{Na}{i} components. However, a number of factors lead us to conclude that the DIBs observed in these sightlines originate in M\,31: 1) The DIB velocities match the peak M\,31 \ion{H}{i} velocities (see Section \ref{sec:HI}), whereas we detect a negligible contribution to the DIB equivalent widths at radial velocities corresponding to \ion{H}{i} in the MW (at $v_{LSR}<45$~\kms; see \citealt{bra09}); 2) LAB survey data towards the targets in question \citep{kalberla05,hartmann97} show a relatively small foreground \ion{H}{i} component, such that the $W(5780)/N({\rm H\, \textsc{i}})$ ratios towards 4412.17 and 4417.80 would be very large if this DIB originated in the MW halo, and inconsistent with the values observed previously in the Milky Way ISM (see Figure 8 of \citealt{mcc10}); 3) Assuming the Galactic halo obeys the average MW DIB $W$ \emph{vs.} \ebv\ relation, the equivalent widths expected from the M\,31 foreground reddening of \ebv~=~$0.06\pm0.02$ are only $32\pm11$ \,m\AA\ for \lam5780 and $71\pm24$\,m\AA\ for \lam6283, whereas the measured DIB equivalent widths towards 4412.17 of $190\pm17$ m\AA\ and $491\pm82$ m\AA, respectively, are consistent with the M\,31 reddening (\ebvm) in this sightline; 4) The \ion{Na}{i} spectrum of 4406.87 in Figure \ref{fig:nai}, which shows no evidence for M\,31 \ion{Na}{i} absorption, gives another indication of the properties of the foreground (MW) neutral gas in Field 3. The quantity and velocity of this foreground gas is very similar to the amount of gas in the other fields where the MW \lam5780 components (which can be easily separated from the M\,31 DIBs), are clearly very weak. Assuming the foreground gas chemical properties are not grossly variable on the scale of the $\sim1$ degree that separates the GMOS fields, this strongly suggests that the foreground DIBs in Field 3 would be similarly as weak. Small-scale structure in the Galactic halo could produce variations in the relative abundances of atoms, dust and DIB carriers, but given the reasons outlined above, it is highly unlikely that the Field 3 DIBs originate in chemically peculiar, compact, high-velocity Galactic clouds.

\subsection{H\,\textsc{i} 21 cm emission and the M\,31 gas-to-dust ratio}\label{sec:HI}

\begin{figure}
\centering
\epsfig{file=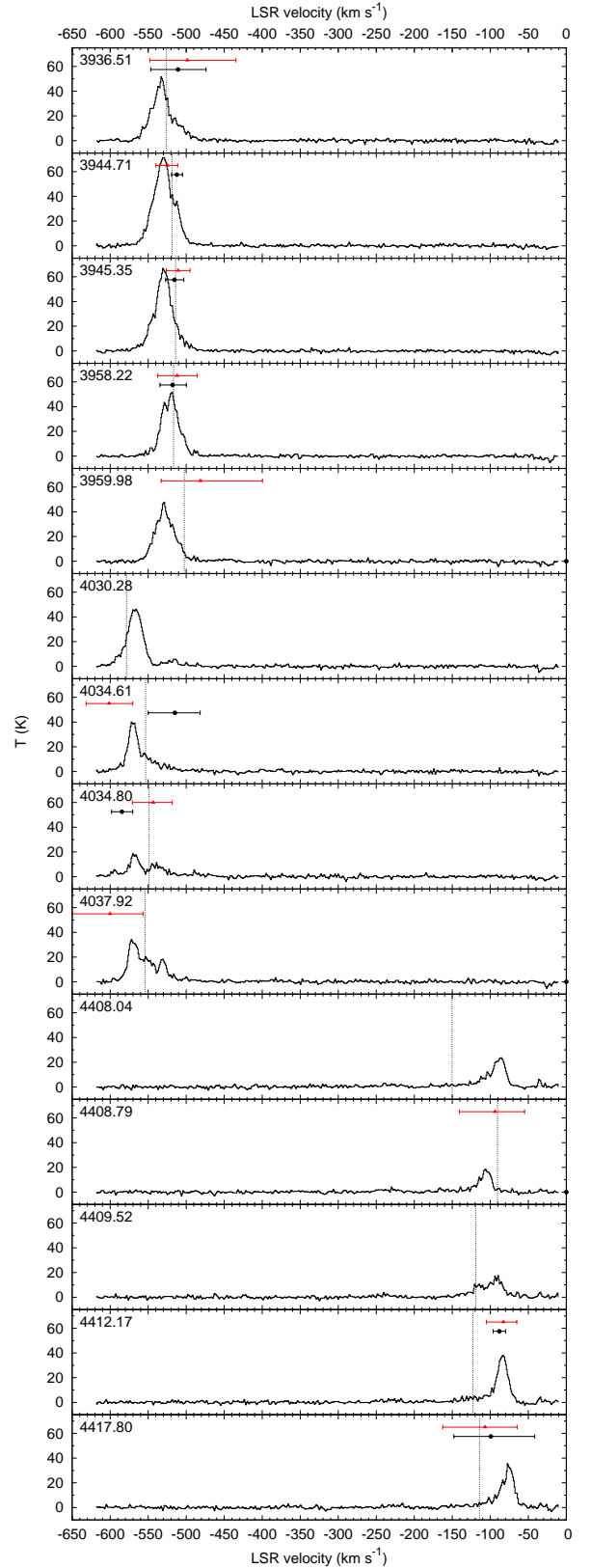, width=\columnwidth}
\caption{\ion{H}{i} emission spectra derived from the survey by \citet{bra09}. M\,31 \ion{Na}{i} velocities are plotted with dotted vertical lines. \lam\lam5780 and 6283 DIB velocities (and $1\sigma$ error bars) are plotted with black circles and red triangles, respectively. }
\label{fig:hi}
\end{figure}

Dr. Robert Braun kindly made available data cubes and opacity-corrected \ion{H}{i} column density maps from the \citet{bra09} 21~cm survey of M\,31, performed using the Westerbork Synthesis Radio Telescope and the Green Bank Telescope. These data are at a spatial resolution of $18''\times15''$, which corresponds to a linear resolution of $70\times56$~pc at the distance of M\,31. The \ion{H}{i} spectra have a velocity resolution of 2.3 \kms. From these data we used bi-linear interpolation to extract opacity-corrected column densities and \ion{H}{i} spectra for our targets (shown in Figure \ref{fig:hi}).

Opacity-corrected \ion{H}{i} column densities and gas-to-dust ratios $G=\frac{1}{2}$\n{\ion{H}{i}}/\ebvm\ are given in~Table \ref{tab:5780}; the gas-to-dust ratio has thus been calculated assuming that half of the total \ion{H}{i} column lies in front of the target stars. However, this fraction is uncertain, so the errors on $G$ are $\pm100$\%.

The average gas-to-dust ratios are 13.4, 8.2 and 7.7  (in units of $10^{21}$ cm$^{-2}$ mag$^{-1}$), for Fields 1-3, respectively, with an overall average of 9.8. Previous measurements of the M\,31 gas-to-dust ratio lie in the range $\approx$\,5--$15\times10^{21}$~cm$^{-2}$\,mag$^{-1}$ \citep{bergh75,bajaja77,kumar79,lequeux00}. These values may be compared with the nearby Galactic value of  $5.8\times10^{21}$~cm$^{-2}$\,mag$^{-1}$ \citep{boh78}, which was derived from \ion{H}{i} absorption lines and therefore does not suffer from the problem of background contamination. It is difficult to draw firm conclusions from these data due to the uncertainties in the fraction of background \ion{H}{i} in M\,31, but is seems reasonable to conclude that the gas-to-dust ratio of M\,31 is broadly similar to that of the Milky Way.  Although many of the observed sightlines appear to have gas-to-dust ratios significantly greater than Galactic, this may be caused by an observational bias towards stars located on the near side of the M\,31 disc; our derived gas-to-dust ratio would be an overestimate in such cases. Accounting for errors, the gas-to-dust ratio is lower towards 4412.17 than the average MW value. This sightline also clearly exhibits relatively weak DIBs per unit reddening.

Fig.~\ref{fig:hi} shows that the velocities of the interstellar clouds as probed using \ion{Na}{i} lines correspond imperfectly with the peak \ion{H}{i} velocity in many cases.  This may be due to the presence of background \ion{H}{i} clouds that do not contribute to the \ion{Na}{i} absorption spectra. However, for three of the Field 3 stars, the \ion{Na}{i} velocities do not match the \ion{H}{i} gas at all, with \ion{Na}{i} velocities at least 20~\kms\ redwards of the \ion{H}{i}.  For 4412.17 and 4417.80, the \ion{Na}{i} velocities are similarly displaced from the DIB velocities (see Section \ref{sec:dibs}), but the DIBs match very closely with the positions of the \ion{H}{i} peak velocities. Given that DIB carriers, \ion{Na}{i} and \ion{H}{i} are correlated \citep[\eg][]{hobbs74,fer85,herbig93,mcc10}, and their velocities are generally closely matched in the diffuse ISM, there is likely to be a problem with the measured M\,31 \ion{Na}{i} cloud velocities in Field 3 (probably for the reasons related to sky-subtraction discussed in Section \ref{sec:nai}). The measured \ion{Na}{i} velocities would be offset to the red (as observed) if the MW component is not as strong as was derived in the \textsc{vapid} model fits or if the sky-line subtraction was incomplete. To resolve this situation would require high resolution \ion{Na}{i} spectra in order to separate the M\,31 and MW components.  The \ion{H}{i} gas in Field 3 spans a relatively narrow range of radial velocities (from about $-120$ to $-60$~\kms), and should be considered as a more reliable indicator of the velocity of the interstellar matter in these sightlines. The resulting velocities from Gaussian fits to the \ion{H}{i} emission profiles are shown for comparison with the M\,31 DIBs in Field 3 in Figures \ref{fig:5780} to \ref{fig:6613}.

CO peak velocities from radio observations by \citet{niet06} are consistent with those of the \ion{H}{i} gas. For Field 3, CO matches better with \ion{H}{i} than \ion{Na}{i}, which provides further evidence that our measured M\,31 \ion{Na}{i} velocities for 4408.04, 4412.17 and 4417.80 are probably in error.

\subsection{M\,31 \ion{H}{ii} regions}\label{sec:hii}

\begin{figure}
\centering
\epsfig{file=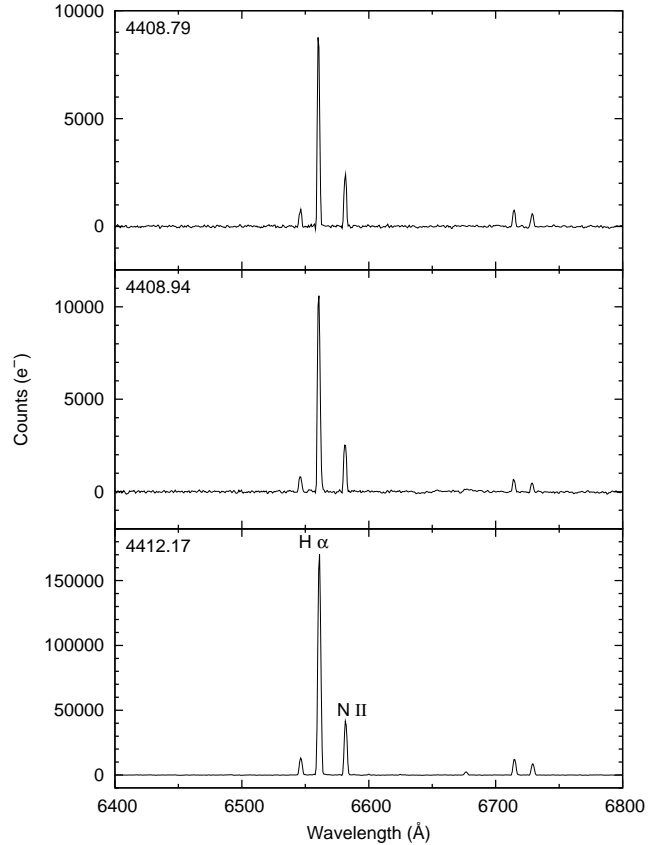, width=\columnwidth}
\caption{Sky-subtracted spectra from observed \ion{H}{ii} emission regions in the telluric rest frame. The \ion{N}{ii}~6583 \AA\ and H~$\alpha$ lines used to derive oxygen abundances are marked.}
\label{fig:hii}
\end{figure}

In addition to the stars in Table \ref{tab:stars}, three \ion{H}{ii} regions were observed serendipitously in Field 3, spatially coincident with the GMOS slits of the stars 4408.79, 4408.94 and 4417.12. Using the relative strengths of the emission lines in these spectra, it is possible to derive an estimate for the interstellar metallicity in the vicinity of Field 3. The usefulness of these spectra is limited because they are not flux calibrated and the blue and red wavelength regions were recorded in separate exposures. Nevertheless, an estimate of the O/H abundance ratio can be obtained from the relative \ion{N}{ii}~6583 \AA\ and H$\alpha$ line fluxes \citep[see][]{pet04}.

The emission spectra were extracted from regions at the edges of the GMOS slits where the stellar flux was negligible, sky-subtracted using sky spectra from nearby slits showing no nebular emission, and continuum-subtracted using low-order polynomial fits. The final spectra are shown in Figure \ref{fig:hii}. The logarithmic flux ratios ($\log$([\ion{N}{ii}]/H$\alpha$)) for 4408.79, 4408.94 and 4417.12 are $-0.54$, $-0.63$ and $-0.61$ respectively, which correspond to oxygen abundances $12 + \log[{\rm O}/{\rm H}]=8.59$, 8.53 and 8.51 (all with $1\sigma$ errors of $\pm0.18$), using equation (2) of \citet{pet04}. The average value, $8.54\pm0.18$, is similar to that in the solar neighbourhood, where the oxygen abundance of \ion{H}{ii} regions is in the range $8.45-8.50$ \citep[see][and references therein]{pil03}.

Our derived [O/H] ratios add to the limited data already available in the literature on the metallicity of M\,31. \citet{trundle02} reviewed previous work on oxygen abundance measurements of \ion{H}{ii} regions in M\,31 and identified that there is some uncertainty in the metallicity gradient across M\,31. Measured oxygen abundance gradients range from $-0.013$ to $-0.027$~dex\,kpc$^{-1}$, depending on the method used to derive [O/H]. Extrapolating these gradients gives $12 + \log[{\rm O}/{\rm H}]=8.70$ to 9.21 at the centre of M\,31, compared with 8.90 in the centre of the Milky Way. The three \ion{H}{ii} regions we observed are approximately $23'$ from the centre of M\,31, which corresponds to 5.2~kpc (assuming a distance of 783 kpc to M\,31; \citealt{holland98}). The average oxygen abundance in the three \ion{H}{ii} regions (8.54), lies towards the lower end of the range of values previously reported at this radius in M\,31.

Fields 1 and 2 lie at radial distances ($r$) of $63'$ ($14.3$~kpc) and $41'$ ($9.3$~kpc) respectively from the centre of M\,31. Using the metallicity relation of $12 + \log[{\rm O}/{\rm H}]=8.7-0.013r$, which was deduced from the study of B star abundances by \citet{trundle02}, we find oxygen abundances of 8.51 and 8.58 for these two fields. Thus, there is no evidence to suggest that the metallicity of the ISM along any of our surveyed sightlines is significantly different from solar.

\subsection{Interstellar radiation field}\label{sec:uv}

Near-ultraviolet interstellar radiation field strengths (NUV ISRFs) were calculated for the M\,31 ISM in each sightline, using the NUV GALEX mosaic of M\,31 (imaged at a central wavelength of 2271 \AA) from \citet{thil05}. The NUV ISRF strengths ($I_{NUV}$) are given in Table \ref{tab:5780} in units of the Galactic NUV ISRF of \citet{draine78}, integrated over the aperture size and NUV wavelength response function of the GALEX telescope. Maps of $I_{NUV}$ for each of our three fields were computed from the GALEX image and are shown in Figure \ref{fig:fields}.

To calculate these $I_{NUV}$ maps, individual pixel fluxes were background-subtracted and corrected for foreground extinction using $A_{NUV}=8.0\times E_{B-V}$ \citep{paz07}, assuming a MW foreground extinction of 0.06~mag (see~Section \ref{sec:fg}). The M\,31 extinction was derived for each field from the average of the \ebvm\ values of the stars listed in Table \ref{tab:stars}. For each sightline in the ISRF maps, $I_{NUV}$ was calculated at a point at a line-of-sight distance $r$ halfway into the foreground M\,31 cloud, using the extinction-distance relation derived by \citet{verg98} for the solar neighbourhood ($dA_V/dr=1.2$~mag\,kpc$^{-1}$, with $R_V=A_V/E_{B-V}=3.1$), and assuming that the observed NUV flux is emitted from the image plane (at a radial distance of 783 kpc). The NUV photons were assumed to be subject to the same level of extinction as a function of distance as in the solar neighbourhood; consequently, the calculated $I_{NUV}$ are dominated by the flux from OB stars in the immediate vicinity of the interstellar clouds of interest; stars at distances greater than 1 kpc make a negligible contribution to the total NUV flux at a given point in space. The NUV ISRF strengths thus calculated are only approximate because they take no account of the fact that the shape of the extinction law, and the extinction as a function of distance in M\,31, may be different from that observed in the Milky Way. The M\,31 foreground extinction is treated as a uniform slab of material, so any small-scale inhomogeneities in the M\,31 ISM are neglected.

Within Fields 1-3 respectively, the interstellar UV field strengths ($I_{NUV}$) averaged over all pixels are 0.24, 0.77 and 0.34 (in units of the Draine field), with an average value of 0.45.  Evidently, even though these fields are relatively rich in OB stars, the NUV ISRF is still significantly weaker than that of the Milky Way. This is consistent with the conclusions of previous authors (see Section \ref{sec:intro}).  Field 2 has the greatest $I_{NUV}$ values due to the presence of OB78 (\object[NGC206]{NGC\,206}) -- the richest OB association in M\,31.  Indeed, the sightlines towards the DIB target stars in this field have NUV field strengths which are comparable to or greater than the \citep{draine78} Galactic value. On the other hand, most of the stars in Field 1 towards which DIBs are observed have $I_{NUV}<0.5$, including the star in our sample 3944.71, towards which the M\,31 DIBs are strongest.

\subsection{Aromatics and dust emission}\label{sec:spitzer}

For the three M\,31 fields presented here, we obtained reduced infrared Spitzer Space Telescope images in the IRAC 8~$\mu$m and MIPS 24~$\mu$m bands from the surveys of \citet{bar06} and \citet{gor06}. The flux in the 8~$\mu$m band is dominated by PAH emission from the 7.7~$\mu$m (and 8.6~$\mu$m) aromatic bands (see, for example \citealt{draine01} and \citealt{flagey06}), whereas the flux in the 24~$\mu$m band predominantly arises from interstellar dust. The ratio of 8~$\mu$m-to-24~$\mu$m fluxes may therefore be used as an approximate measure of the relative PAH emission intensity per unit dust mass \citep[\eg][]{flagey06,engelbracht08,gordon08}, and as such, provides an indication of the abundance and excitation level of PAHs in the ISM.

The 8~$\mu$m IRAC images are shown for our three fields in Figure \ref{fig:fields}. The interstellar emission flux was measured for each of the sightlines in Table \ref{tab:5780} from the background-subtracted Spitzer images using a ring median filter of radius 5 pixels and width 2 pixels in order to exclude the flux from stellar point-sources. To measure the 8/24~$\mu$m interstellar flux ratio, the 8~$\mu$m images were convolved and rebinned to the lower spatial resolution of the 24~$\mu$m images using the convolution kernel provided by \citet{gordon08}.

In the plane of the Milky Way, the interstellar 8/24~$\mu$m flux ratio ($R$) varies between about 0.1 and 2.0, with a mean value of 1.6 \citep{flagey07}. Similarly, in nearby spiral galaxies, $R$ also varies between about 0.1 and 2, with considerable scatter within individual galaxies \citep{bendo08}. The ratio $R$ correlates with metallicity, which has been interpreted as being due to variation in the strength of PAH emission relative to the 24~$\mu$m dust emission, at least partly as a result of differences in PAH abundances \citep{eng05}.  However, PAH emission correlates better with ionization of the ISM than the metallicity \citep{gordon08,bendo08}, so the 8~$\mu$m flux may be a better indicator of the degree to which the PAHs are ionised rather than their overall abundance. $R$ also tends to be lower inside \ion{H}{ii} regions and higher in denser, dustier gas associated with PDR boundaries.

In our GMOS fields, $R$ values range between 0.63 and 1.78, with mean values of 1.05, 1.16 and 1.04 in Fields 1-3, respectively.  Thus, there is nothing particularly unusual about the PAH emission in these fields compared with other galaxies, although the mean values are somewhat lower than observed in the Milky Way. This is perhaps indicative of a reduced level of ionisation or processing of the aromatic material in our M\,31 fields as a result of the lower UV radiation fields that are typical of M\,31.

The strength of the 8~$\mu$m emission discussed here is at variance with previous studies \citep{cesarsky98,pagani99} that identified anomalously weak aromatic mid-IR emission bands in the bulge and nucleus of M\,31.  Thus, the PAH emission traced by the 8~$\mu$m flux, at least in our observed fields (which are away from the nucleus/bulge), is apparently not as weak as previously thought, and implies that the PAHs in M\,31 may in fact not be dissimilar to those found in the Milky Way and other spiral galaxies. It must be noted, however, that the NUV ISRF in our observed fields may be greater than is typical for M\,31 due to the concentration of OB stars, which could give rise to increased 8~$\mu$m emission in these regions relative to the average conditions in M\,31.

\section{DIB strengths and correlations}

\begin{figure}
\centering
\epsfig{file=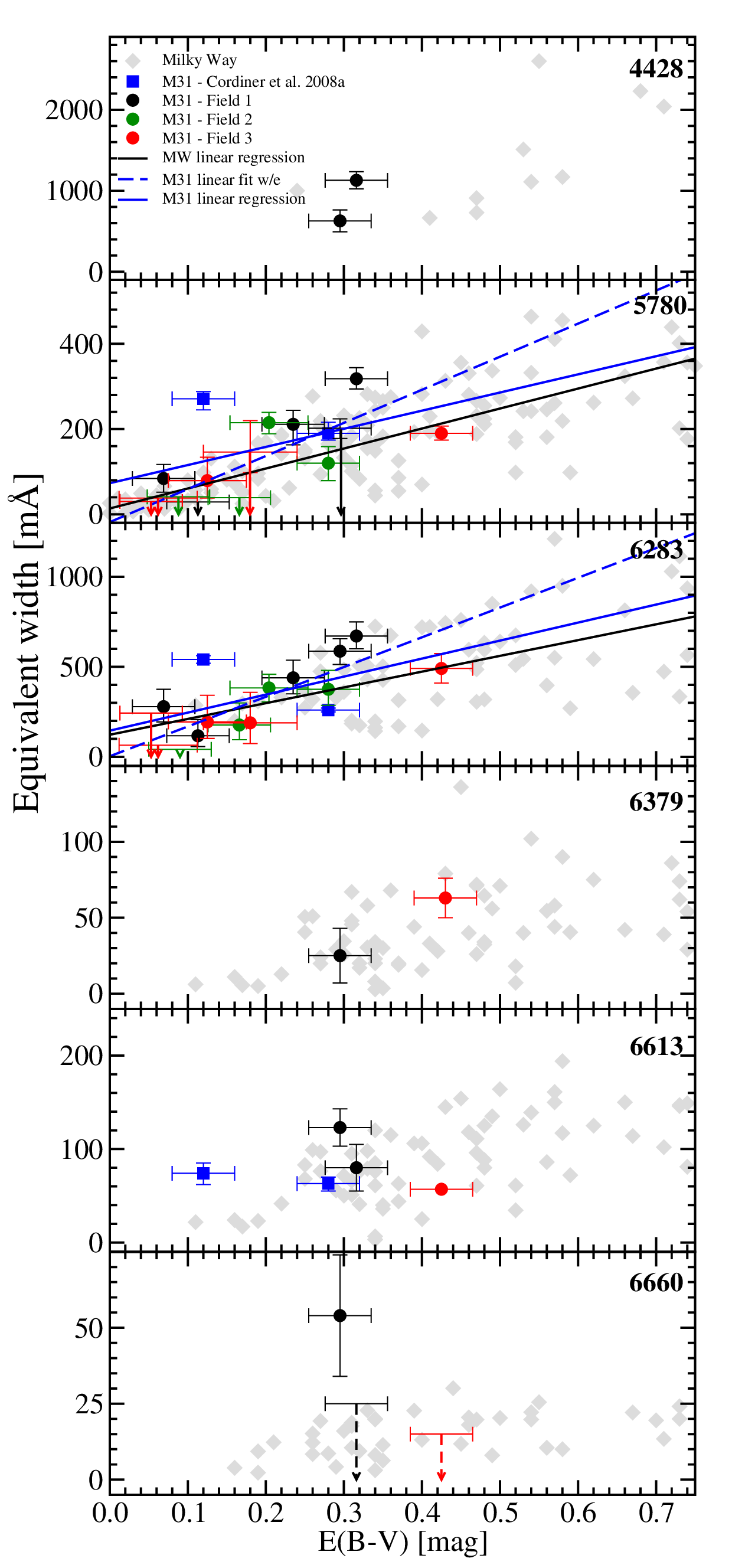, width=\columnwidth}
\caption{\lam\lam4428, 5780, 6283, 6379, 6613 and 6660 DIB equivalent widths as a function of \ebv\ for M\,31 and the Milky Way. Data points for the three GMOS fields are plotted in different colours. Measurements from \citet{cor08} are included (blue squares). DIB equivalent width upper limits are shown with downward-pointing arrows. MW DIB data are from \citet{herbig93}, \citet{thorb03}, \citet{megier05} and \citet{cord06}, and for \lam4430, \citet{herb75}. For the \lam\lam 5780 and 6283 DIBs, linear regressions are shown for the MW data (black solid lines) and the M\,31 data (blue solid lines). A linear fit that takes into account the errors in both ordinates is also shown for the M\,31 \lam\lam5780 and 6283 data (dashed blue lines).}
\label{fig:DIBCorrelationEbv}
\end{figure}

\begin{figure*}
\begin{center}
\includegraphics[angle=0,width=0.95\columnwidth]{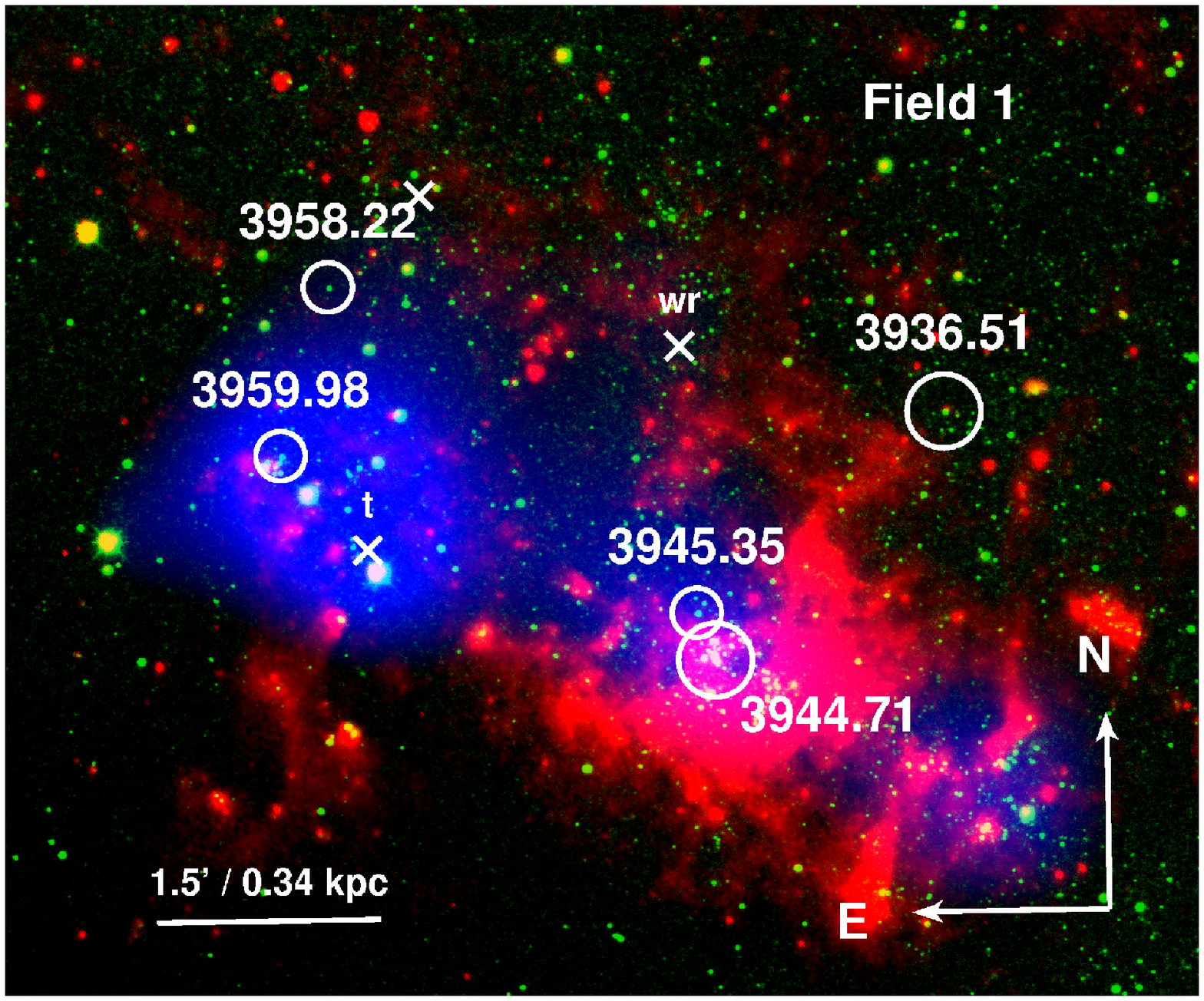}
\includegraphics[angle=0,width=0.95\columnwidth]{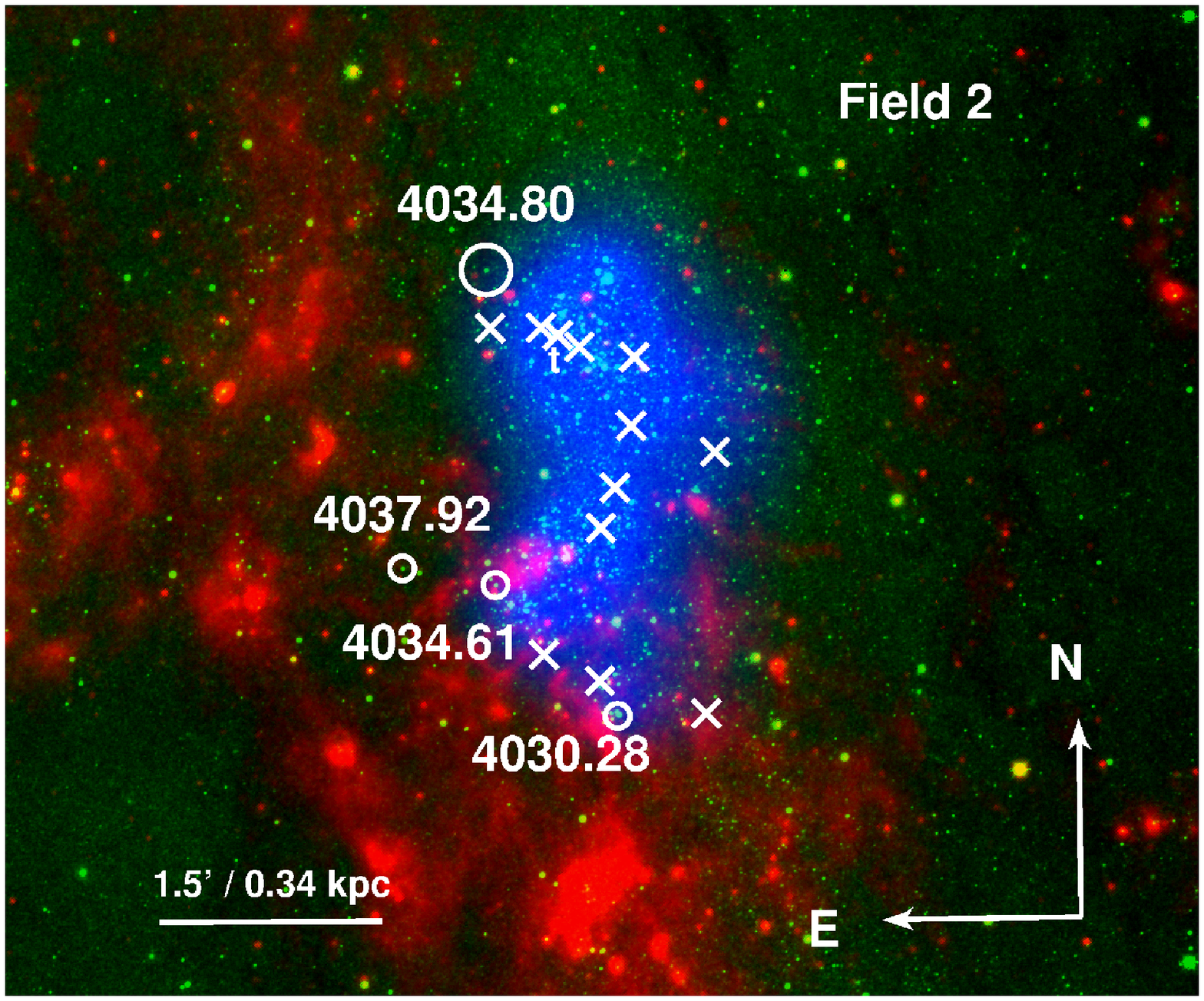}
\includegraphics[angle=0,width=0.95\columnwidth]{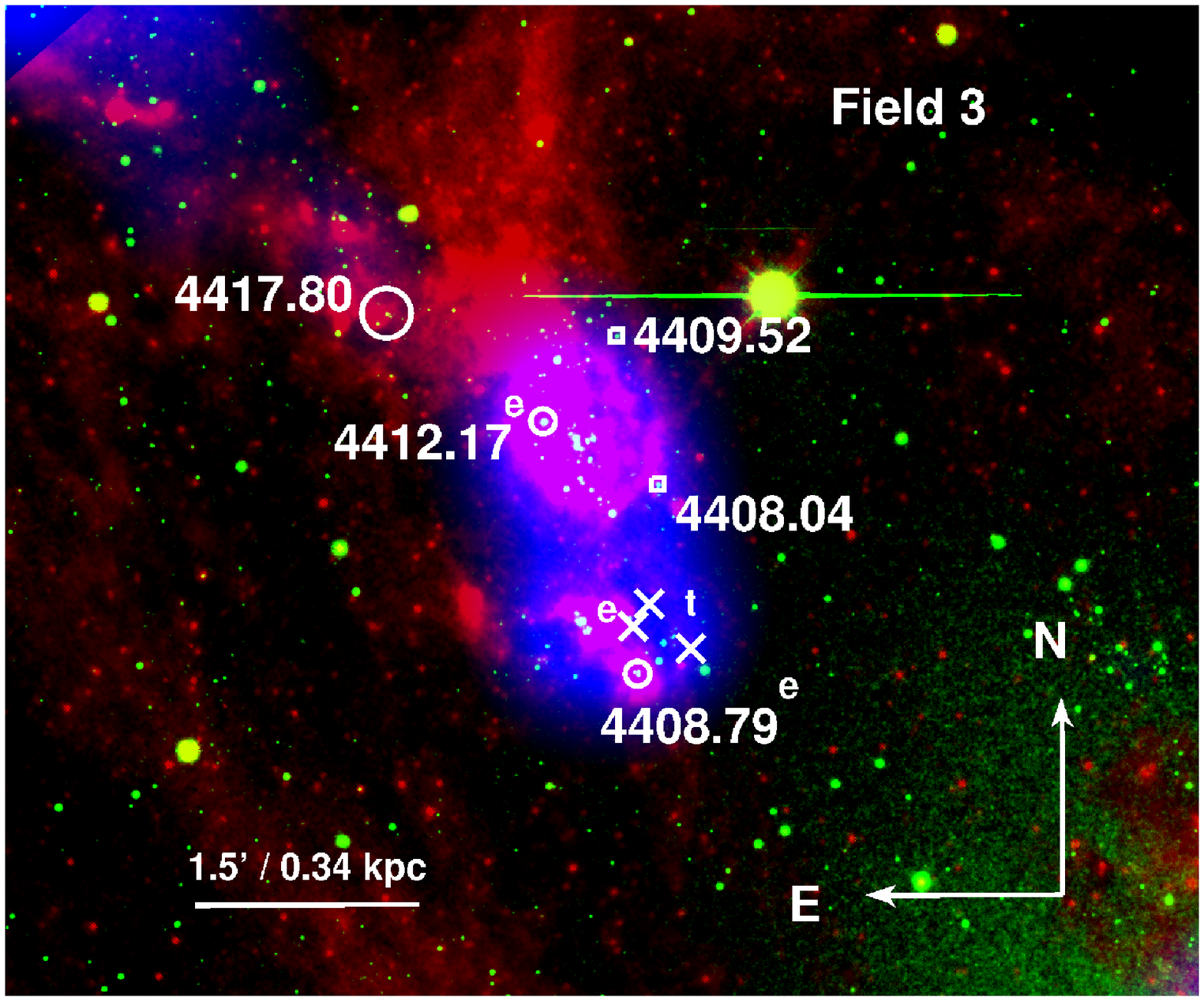}
\end{center}
\caption{Three-colour composite maps of our observed M\,31 fields. Blue corresponds to the NUV ISRF (see Sect.~\ref{sec:uv}), green is the optical $V$-band image (\citealt{massey06}), and red is the Spitzer IRAC 8~$\mu$m data that traces PAH emission (Sect.~\ref{sec:spitzer}).
Intensities are shown on a logarithmic scale. ISRF values range from 0.05 to 0.5, 6.0 and 1.0 Draine for Fields 1-3, respectively. 8~$\mu$m flux values range from 0.0 to 50.0 MJy\,sr$^{-1}$ for all fields. Lines-of-sight with M\,31 DIBs are marked with circles, the sizes of which denote the strength of the \lam6283 DIB per unit reddening (relative to the mean M\,31 DIB strength): small, 10\arcsec-circles denote weak; medium, 20\arcsec-circles denote average and large, 30\arcsec-circles denote strong DIBs. Crosses indicate sightlines with no DIB detections, squares denote upper limits. Also marked are the telluric standards (t), the \ion{H}{ii} emission regions (e) and the single Wolf-Rayet star observed in Field 1 (wr). The two targets with previously observed DIBs reported by \citet{cor08} are about 6 arcminutes ($\approx1.3$~kpc) south of Field~2.}
\label{fig:fields}
\end{figure*}

Figure \ref{fig:DIBCorrelationEbv} shows the relationships between the equivalent widths of a selection of the strongest diffuse interstellar bands (\lam\lam4428, 5780, 6283, 6379, 6613 and 6660) and interstellar reddening (\ebvm), for sightlines in M\,31 and the Milky Way. All of the newly-observed M\,31 DIB measurements lie within the range of values previously observed in the Milky Way, apart from \lam6660, which seems anomalously strong, perhaps due to spectral contamination, which may be indicated by reference to the telluric standard spectrum in Figure \ref{fig:6660}.

The majority of the M\,31 data points plotted in Figure \ref{fig:DIBCorrelationEbv} lie within the middle-to-upper part of the range of values covered by the MW data.  Thus, per unit \ebv, the M\,31 DIBs are as strong or stronger than those typically observed in the Milky Way. The relative enhancement of DIB strengths is most pronounced for Field 1 (shown as black filled circles in the plots), for which the equivalent widths of the strongest (and therefore most reliably measured) DIBs \lam\lam5780, 6283 and 6613 are clearly at the upper end of the range of Galactic values. The locations of those sightlines with the strongest DIBs are shown in Figure \ref{fig:fields}; the sizes of the circles indicate the relative strength of the \lam6283 DIB per unit \ebv\ in each sightline. It is evident that the ISM in Field 1 hosts the strongest DIBs per unit \ebv\, and Fields 2 and 3 the weakest.  The DIBs with the largest equivalent widths in our sample are towards 3944.71, 3945.35 in Field 1. These are the sightlines with the most reddening and the greatest \ion{H}{i} column densities. Relative to the amount of dust in the 3944.71 sightline, the \lam\lam5780, 6283 and 6613 equivalent widths are at the very upper limit of what has previously been observed in the Milky Way, so these sightlines have exceptionally strong DIBs. As can be seen in Table \ref{tab:dibs}, the majority of the measured DIBs in the sightlines towards 3944.71 and 3945.35 are significantly stronger than those in the Milky Way, for the same reddening. These two sightlines have relatively low NUV ISRF strengths -- about 0.5 and 0.4 times that of the Milky Way, respectively.

As derived in Section \ref{sec:fg}, the mean \lam5780 equivalent width per unit reddening measured in the Milky Way is $[W(5780)/E_{B-V}]_{mean}=0.53\pm0.23$~\AA\,mag$^{-1}$. Using data on the \lam6283 DIB from the same Galactic literature, $[W(6283)/E_{B-V}]_{mean}=1.18\pm0.53$~\AA\,mag$^{-1}$. The errors are the standard deviations of the respective data. The mean values for $W(5780)/E_{B-V}$ in Fields 1, 2 and 3 are 0.93, 0.7 and 0.73~\AA\,mag$^{-1}$, respectively, and for $W(6283)/E_{B-V}$, 2.4, 1.37 and 1.27. These values are consistent with the identified trend that our observed M\,31 DIBs are relatively strong compared with the Milky Way, and highlight in particular, the strengths of these DIBs in Field 1.

\begin{figure*}
\begin{center}
\epsfig{file=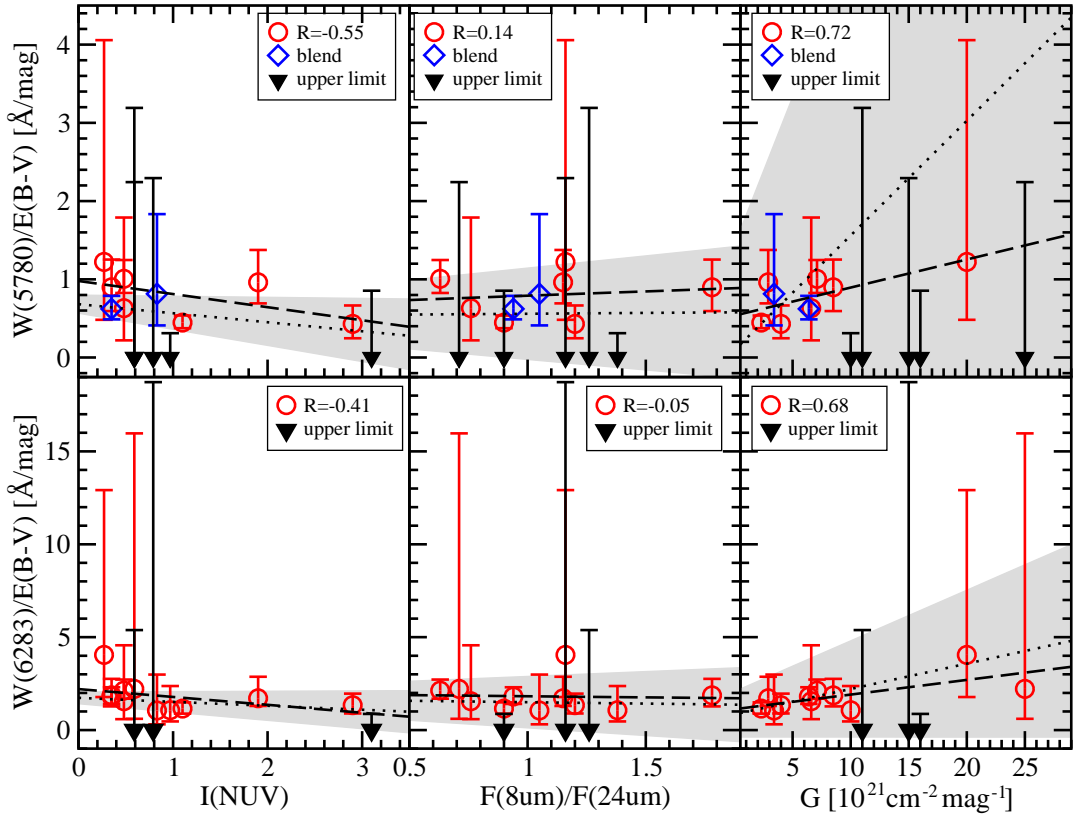, angle=270, width=1.8\columnwidth}
\end{center}
\caption{Correlations between \lam\lam5780 and 6283 DIB equivalent widths per unit \ebv\ and interstellar physical / chemical properties: Interstellar UV field strength ($I_{NUV}$), Spitzer 8/24~$\mu$m flux, and logarithmic gas-to-dust ratio ($G$). Linear regressions have been plotted (dashed lines) and correlation coefficients $R$ are displayed in the top right. Fits that take into account the error bars are shown with dotted lines, the 1$\sigma$ ranges for which are shown with grey polygons. Fitting excludes blended DIBs and upper limits.}
\label{fig:correl}
\end{figure*}

The weakest DIBs (per unit \ebv) are observed towards 4030.28, 4034.61 and 4037.92 in Field 2 and 4412.17 in Field 3.  These four sightlines are near to regions of strong UV flux (see Figure \ref{fig:fields}), and include the strongest NUV ISRFs of the observed sightlines, with $I_{NUV}\gtrsim1$. The smallest $W(6283)/E_{B-V}$ value is observed towards 4030.28, where $I_{NUV}$ is largest, and the largest $W(6283)/E_{B-V}$ value is observed towards 3936.51, where $I_{NUV}$ is smallest, which is suggestive of a correlation between the interstellar UV field and DIB strengths. To analyse this correlation, the equivalent widths per unit reddening of the two strongest DIBs (with the smallest relative measurement errors), \lam\lam 5780 and 6283, are plotted with respect to the NUV ISRF strength in Figure \ref{fig:correl}. There is a significant negative correlation between $W$/\ebvm\ and $I_{NUV}$ for both DIBs at the 68\% confidence level, shown by the grey polygon.  The correlation coefficients between the data for \lam\lam5780 and 6283 are $-0.55$ and $-0.41$, respectively. This indicates that as the interstellar UV flux increases, the abundances of both of these DIB carriers tend to become less, relative to the amount of dust. Plausible explanations for this phenomenon are that the DIB carriers become ionised or are photodissociated, or that the chemical environments in which the carriers form are destroyed by NUV radiation.  

A link between UV flux and DIB strengths is corroborated by the pattern of DIB observations in Field 2 and the morphology of the ISM in this region.  This field covers the OB78 association, which is the largest, most active region of recent star formation in M\,31.  Most of the stars at the heart of this association do not show significant reddening or DIBs (see Figure \ref{fig:fields}) -- the majority of the gas and dust around these stars has been blown away by strong stellar winds and supernovae \citep{brinks84,loi96}, creating a large, hollow region in the ISM several hundred parsecs across.  There is still sufficient interstellar matter at the periphery of the central cluster that DIBs are observed towards many of the nearby stars, and the 8~$\mu$m flux is intensely strong as a result of the high UV-optical flux that excites the interstellar aromatics. The DIBs are generally quite weak in this field compared to the M\,31 average, which we suggest is due to the destruction by strong UV irradiation of the environments in which the carriers form.

The relationship between DIB strengths and UV flux has been examined previously in the Milky Way and in the Magellanic Clouds. Indeed, the relative weakness of the DIBs observed in the Magellanic Clouds may be attributable to the strong radiation fields that pervade these galaxies \citep{cox06,cord06,welty06}. UV-rich regions such as those found in the vicinity of the Orion Nebula and Scorpius are also known to show weak DIBs \citep{cami97,snow95}. \citet{welty06} found no significant correlation between UV flux and DIB strengths in the MCs. However, the UV field strengths in their analysis were derived from H$_2$ excitation, and it is known that the DIB carriers do not predominantly reside in the relatively dense regions of the ISM traced by H$_2$ \citep[\eg][]{herbig93}.

The narrow (\lam\lam5797, 6613, 6660 and 6993) DIBs are enhanced towards 3945.35 with respect to the average MW diffuse ISM, while the broad DIBs are not. On the other hand, the line of sight towards 3944.71 (where the DIBs are strongest), shows enhanced broad ($\lambda\lambda$5780, 6203 and 6283) DIBs with respect to the average MW value. Previous studies in the Milky Way \citep[\eg][]{cami97,krel98}, identified that these broad DIBs tend to be strong in relatively strongly UV-irradiated sightlines (where the narrow DIBs are weak), whereas narrow DIBs are strong in well-shielded sightlines.  However, the negative correlation we have identified between the strengths of the \lam\lam5780 and 6283 DIBs and $I_{NUV}$ suggests that the environments or the carriers of these DIBs are destroyed by UV radiation. Their large strengths in relatively strongly-irradiated Galactic sightlines seems to be at odds with our observed correlation, but could instead be evidence for the production of enhanced quantities of DIB carriers as a result of photochemistry or dust photo-abrasion in these Galactic regions. Indeed, it is notable from Figure \ref{fig:fields} that the two sightlines with the strongest DIBs (3944.71 and 3945.35), reside in a region of both strong 8~$\mu$m flux (indicative of dense, dusty gas) and moderately high UV field strength.  Given the association of mid-IR emission with ionised PAHs \citep{tie08}, this suggests that the carriers of the \lam\lam5780, 6203 and 6283 diffuse bands may well be ionised molecules.

The correlations between \lam\lam5780 and 6283 and \ebv\ are shown in Figure \ref{fig:DIBCorrelationEbv}. Taking into account the error bars on $W$ and \ebv, least-squares linear fits to the data result in the parameters shown in Table \ref{tab:regression}. Within the quoted $1\sigma$ errors, there are no significant differences between the slopes and intercepts of M\,31 and MW data for either DIB. Kolmogorov-Smirnov (K-S) tests were applied to the residuals of the observed data compared with the MW fits. The resulting probabilities $(1-P)$ describe the likelihood that the M\,31 and MW DIB data have different distributions with respect to \ebv. For \lam5780, $(1-P)=0.93$, which shows there is a high statistical probability that the M\,31 and MW DIBs follow a different relationship with \ebv. This is consistent with our observation that the M\,31 DIBs are typically rather strong compared with in the Milky Way. For \lam6283, the derived value of $(1-P)=0.50$ is inconclusive, and indicates that more observational data are needed.

\begin{deluxetable}{lccc}
\tablewidth{0pt}
\tablecaption{Fit parameters for M\,31 \lam\lam5780 and 6283 DIB equivalent widths \emph{vs.} reddening \label{tab:regression}}
\tablehead{
DIB&intercept $a$&slope $b$&correlation coefficient $R$
}
\startdata
MW 5780&14 $\pm$ 10&469 $\pm$ 29 & 0.81 \\    
M\,31 5780&73 $\pm$ 62& 426 $\pm$ 234 & 0.60 \\
MW 6283&123 $\pm$ 80  & 876 $\pm$ 174 & 0.55\\
M\,31 6283&146 $\pm$ 102 & 1000 $\pm$ 430 & 0.57\\
\enddata
\tablecomments{Equivalent width $W = a +(b \times E_{B-V})$. The outlier data from MAG 70817 \citep{cor08} were omitted from the fits.}
\end{deluxetable}

Relatively weak DIBs are observed towards 4412.17 which has a very low $G$ value; relatively strong DIBs are observed towards 3936.51, which has a high $G$ value.  These results could be explained if variations in $G$ are caused by the destruction of dust grains (by shocks and UV radiation), and if the grain degradation products contain or give rise to the DIB carriers. Conversely, if gas-phase species are depleted out onto the dust, reducing the gas-to-dust ratio, then molecular DIB carriers would also accrete onto the grains, making the bands weaker. However, when accounting for the substantial errors on $G$ across all sightlines, there is no statistically significant correlation between the gas-to-dust ratio and $W(5780)$ or $W(6283)$, as shown by the grey polygons in Figure \ref{fig:correl}. We know of no prior reports of correlations between DIB strengths and $G$ in the MW; in the LMC, $W$/\ebv\ and $G$ have been shown to be uncorrelated, albeit with a very limited number of data points \citep{cox06,cord06}. Thus, an alternative explanation may be required for the weakness of the DIBs towards 4412.17, perhaps related to the relatively high strength of the 8~$\mu$m emission at the position of this star. This indicates the presence of dense, dusty gas of the kind where DIBs tend to be relatively weak in the Milky Way \citep{snow95}.

In Section \ref{sec:spitzer}, the relationship between the mid-IR 8/24 micron Spitzer flux ratios and the abundance of PAHs was discussed.  Given that some of the DIBs are hypothesised to be caused by electronic transitions of neutral and ionised PAHs (the same classes of molecules believed to cause the 8~$\mu$m emission), it is of interest to look for a correlation between the DIB strengths and the mid-IR flux ratio.  However, as shown by Figure \ref{fig:correl}, there is no relation evident between the strengths of $\lambda\lambda$5780 \& 6283 per unit \ebv\ and the 8/24 micron ratios observed in our M\,31 sightlines. This indicates that the carriers of these particular DIBs are not the same as the aromatics which are predominantly responsible for the 8~$\mu$m emission. There is some uncertainty in this result due to the fact that -- in common with the \ion{H}{i} 21~cm emission -- it is impossible to separate the background contributions to the IR flux in our observed sightlines, so no definitive conclusions may be drawn. Nevertheless, it is apparent from the maps in Figure \ref{fig:fields} that the distributions of sightlines with strong or weak DIBs do not generally follow the 8~$\mu$m emission intensity.

The exceptionally strong DIBs that were previously detected by \citet{cor08} towards MAG~70817 in OB78 now seem to be unusual compared with typical M\,31 and MW trends. The strong \lam5780 DIBs reported in the more extensive Keck DEIMOS study of OB78 by \citet{cox08b} were known to have large error bars on $W$ and \ebv, and the deduction of that study, that the M\,31 DIBs are substantially stronger than those observed in the Milky Way, is not supported by the more accurate and statistically meaningful results presented here.

Past studies, highlighted in Section \ref{sec:intro}, have discovered different dust extinction and polarisation properties in M\,31 compared to the MW, and in particular, a lack of small graphitic dust grains and an abundance of (relatively less UV-processed) PAHs. The similarity of the DIB spectrum to that in the MW indicates that the cause of these differences in the dust properties has little impact on the DIB carriers.  That markedly different UV-extinction and optical polarisation curves have been observed in M\,31 adds to the body of evidence that DIB carriers are not closely related to larger interstellar dust particles or to the smaller graphitic grains believed to be responsible for the 2175 \AA\ UV extinction bump. However, studies of M\,31 dust and extinction have so far been very limited, and little is presently known about the detailed characteristics of the average M\,31 extinction curve. Further dedicated studies of this are warranted.

The general lack of UV/FUV radiation throughout M\,31 compared with the Milky Way (see Section \ref{sec:intro}), due to the relatively low numbers of young stars, apparently does not have a detrimental effect on the DIB carrier abundances.  Thus it seems that the processing of interstellar material by strong, hard UV radiation may not be required to produce DIB carriers.  There is still, however, a significant NUV flux throughout much of M\,31 (which, in the vicinity of dense clusters of OB stars, matches or exceeds the MW average), so the requirement for lower-energy UV photons in the production of DIB carriers cannot be ruled out.

\section{Conclusion}

We obtained optical spectra of over 30 early-type stars in three fields in M\,31 using the Gemini Multi-Object Spectrograph.  Spectral classifications were presented, comprising the first spectral types for 20 of these stars. Reddenings (\ebv) towards our targets were calculated, including the foreground-corrected M\,31 component. The M\,31 and MW components of interstellar \ion{Na}{i} towards the most-reddened subset of our targets were measured and their their column densities, Doppler widths and peak radial velocities derived.

Eleven diffuse interstellar bands (\lam\lam4428, 5705, 5780, 5797, 6203, 6269, 6283, 6379, 6613, 6660, and 6993) have been detected in the M\,31 ISM, with profiles and patterns of strengths similar to those found in the Milky Way and with radial velocities generally matching the interstellar \ion{Na}{i} and \ion{H}{i} and the stellar velocities. Measured radial velocities of the stars and interstellar matter match those of previously observed emission-line objects at the same projected galactocentric radii in M\,31.

The M\,31 \lam\lam5780 and 6283 DIB equivalent widths as a function of \ebv\ fall within the envelope of values typically seen in the Milky Way. However, contrary to LMC and SMC DIBs which are weaker than those in the Galaxy, the M\,31 sightlines generally have DIBs that are as strong or slightly stronger than the average Galactic values. The M\,31 and MW DIB $W$ \emph{vs.} \ebv\ distributions were compared using a K-S test, which yielded a 93\% probability that the observed M\,31 \lam5780 DIB data are different from that of the MW. However, this statistic does not account for the errors in the data values. We measure a mean M\,31 \lam5780 DIB equivalent width per unit reddening of $[W(5780)/E_{B-V}]_{mean}=0.78^{+0.72}_{-0.30}$, compared with $0.53\pm0.23$~\AA\,mag$^{-1}$ in the MW. Similarly, $[W(6283)/E_{B-V}]_{mean}=1.81^{+2.98}_{-0.81}$~\AA\,mag$^{-1}$, compared with the MW value $1.18\pm0.53$~\AA\,mag$^{-1}$. Thus, although the observed mean values are indicative of stronger-than-average DIBs within M\,31, the sizes of the error bars preclude the identification of any significant differences between the DIB strengths in M\,31 and the Milky Way.  Higher signal-to-noise observations of more M\,31 sightlines are required to resolve this issue.

A correlation between DIB strengths and $I_{NUV}$ has been found at a 68\% confidence level, which indicates that the DIB carrier abundances are enhanced in regions with weaker UV interstellar photon flux.  We therefore hypothesise that the general lack of UV photons in parts of M\,31 benefits the survival of DIB carriers and their chemical precursors against photo-destruction, and that FUV radiation is not a requirement for the production of DIB carriers.

The DIBs observed in the spectrum of 3944.71 are very strong per unit \ebv. It is clear, however, that the unusually strong DIBs previously detected in M\,31 by \citet{cor08} and \citet{cox08b} are anomalous with respect to the average M\,31 trend, and that in general, the relationship between M\,31 DIB equivalent widths and reddening is very similar to that of the Milky Way. The degree of scatter of the M\,31 DIBs around the average relationship (per unit \ebv), is also similar to that observed in the Galaxy.

The similarity of the M\,31 DIB $W$/\ebv\ relation to that of the Milky Way suggests that DIB equivalent widths may be applicable as probes of \ebv\ in other, similar spiral galaxies. However, there is evidence that the relationship for M\,31 is slightly different, perhaps as a result of the low UV field strength, and therefore further study of DIBs in external galaxies is warranted. Particular caution must be exercised when using DIBs as a measure of reddening in galaxies with substantially different metallicities, gas-to-dust ratios and interstellar radiation field strengths (such as the Magellanic Clouds).

When taking into account the substantial errors on $G$, the strengths of the \lam\lam5780 and 6283 DIBs (measured by $W$/\ebv) do not show any significant correlation with the gas-to-dust ratio $G$.

No significant correlation is found between $W$/\ebv\ and the 8/24~$\mu$m flux ratio for the \lam\lam5780 and 6283 DIBs, indicating that the (PAH) carriers of the $8$~$\mu$m infrared band are not closely related to these DIB carriers. The 8/24~$\mu$m and gas-to-dust ratios in our fields are in the range of values previously observed in the Milky Way, so in these respects the gas and dust properties of M\,31 are not found to be distinguishably different from those of the Galaxy.

Finally, it should be noted that our results reflect the behaviour of the ISM in only three small regions of M\,31 and therefore do not necessarily reflect the global behaviour of the M\,31 disc. However, our regions do span a range of galactocentric distances and have a range of different properties in terms of the structure and composition of the gas, the density and type of stellar objects, the radiation field strength and the metallicity, and are therefore indicative of the DIB behaviour expected throughout much of the rest of the disc.

\acknowledgments

This article is based on observations (GN-2007B-Q-116) obtained at the Gemini Observatory, which is operated by the Association of Universities for Research in Astronomy, Inc., under a cooperative agreement with the NSF on behalf of the Gemini partnership. For financial support, MAC acknowledges the NASA Institute for Astrobiology, NLJC acknowledges the Faculty of the European Space Astronomy Centre (ESAC) and KTS acknowledges the Engineering and Physical Sciences Research Council (EPSRC). MAC and KTS thank ESAC for visitor funding. We thank Dr. Fabio Bresolin for discussions regarding the \ion{H}{ii} region spectra and metallicities, Prof. Paul Crowther for his classification of 3945.82, and Dr. Ian Hunter for his spectral synthesis calculations. We gratefully acknowledge Prof. Elias Brinks and Dr. Robert Braun for providing their M\,31 21~cm data. This research has made use of the SIMBAD database (operated at CDS, Strasbourg, France), and SAOImage DS9 (developed by Smithsonian Astrophysical Observatory).

{\it Facilities:} \facility{Gemini North (GMOS)}; \facility{Spitzer Space Telescope}.

\bibliographystyle{apj}
\bibliography{M31refs}

\end{document}